\documentclass[aps,prd,floatfix,reprint]{revtex4-2}
\pdfoutput=1
\usepackage[colorlinks=true,linkcolor=blue,citecolor=blue]{hyperref}
\usepackage{graphicx}
\usepackage{bm}
\usepackage{amsmath}
\usepackage{amssymb}
\usepackage{epstopdf}
\usepackage{comment}
\usepackage{xcolor}
\usepackage{multirow}

\newcommand{\orcidicon}{\includegraphics[width=0.32cm]{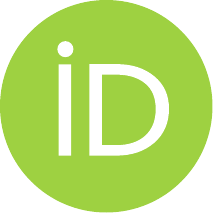}}
\newcommand{\orc}[1]{\href{https://orcid.org/#1}{\orcidicon}}

\newcommand{\orcA}{0000-0001-8217-1484}
\newcommand{\orcB}{0000-0001-5038-8427}
\newcommand{\orcC}{0000-0001-5474-2649}

\newcommand*{\keV}{\text{ keV}}
\newcommand*{\eV}{\text{ eV}}

\newcommand*{\beqn}{\begin{equation}}
\newcommand*{\eeqn}{\end{equation}}
\newcommand{\req}[1]{Eq.\,(\ref{#1})}
\newcommand{\rf}[1]{Fig.~{\ref{#1}}}

\newcommand{\rsec}[1]{Sect.\,{\ref{#1}}}

\newcommand\Tstrut{\rule{0pt}{2.6ex}} 
\newcommand\Bstrut{\rule[-0.9ex]{0pt}{0pt}} 
\newcommand{\TBstrut}{\Tstrut\Bstrut} 

\begin{document}

\title{Matter-antimatter origin of cosmic magnetism}
\author{Andrew Steinmetz\orc{\orcC}}
\author{Cheng Tao Yang\orc{\orcB}}
\author{Johann Rafelski\orc{\orcA}}
\affiliation{Department of Physics, The University of Arizona, Tucson, AZ 85721, USA}

\date{August 22, 2023}

\begin{abstract}
We explore the hypothesis that the abundant presence of relativistic antimatter (positrons) in the primordial universe is the source of the intergalactic magnetic fields we observe in the universe today. We evaluate both Landau diamagnetic and magnetic dipole moment paramagnetic properties of the very dense primordial electron-positron $e^{+}e^{-}$-plasma, and obtain in quantitative terms the relatively small magnitude of the $e^{+}e^{-}$ magnetic moment polarization asymmetry required to produce a consistent self-magnetization in the universe.
\end{abstract}


\maketitle

\section{Introduction}
\label{sec:introduction}
\noindent Macroscopic domains of magnetic fields have been found around compact objects (stars, planets, etc.); between stars; within galaxies; between galaxies in clusters; and in deep extra-galactic void spaces. The bounds for intergalactic magnetic fields (IGMF)  at a length scale of $1{\rm\ Mpc}$ are today~\cite{Neronov:2010gir,Taylor:2011bn,Pshirkov:2015tua,Jedamzik:2018itu,Vernstrom:2021hru}
\begin{align}
 \label{igmf}
 10^{-8}{\rm\ G}>\mathcal{B}_{\rm IGMF}>10^{-16}{\rm\ G}\,.
\end{align}
Considering the ubiquity of magnetic fields in the universe~\cite{Giovannini:2017rbc,Giovannini:2003yn,Kronberg:1993vk}, we search for a common cosmic primordial mechanism considering the electron-positron $e^{+}e^{-}$-pair plasma~\cite{Rafelski:2023emw,Grayson:2023flr}: We investigate the novel hypothesis that the observed IGMF originates in the large scale non-Amp\'erian (i.e non-current sourced in the `Gilbertian' sense~\cite{Rafelski:2017hce}) primordial magnetic fields (PMF) created in the dense cosmic $e^{+}e^{-}$-pair plasma by magnetic dipole moment paramagnetism competing with Landau's diamagnetism. 

Our study of pre-recombination Gilbertian dipole moment magnetization of the $e^{+}e^{-}$-plasma is also motivated by the difficulty in generating Amp\'erian PMFs with large coherent length scales implied by the IGMF~\cite{Giovannini:2022rrl}, though currently the length scale for PMFs are not well constrained either~\cite{AlvesBatista:2021sln}. The conventional elaboration of the origins for cosmic PMFs are detailed in~\cite{Gaensler:2004gk,Durrer:2013pga,AlvesBatista:2021sln}.

Faraday rotation from distant radio active galaxy nuclei (AGN)~\cite{Pomakov:2022cem} suggest that neither dynamo nor astrophysical processes would sufficiently account for the presence of magnetic fields in the universe today if the IGMF strength was around the upper bound of ${\cal B}_{\rm IGMF}\simeq30-60{\rm\ nG}$ as found in Ref.~\cite{Vernstrom:2021hru}. The presence of magnetic fields of this magnitude would then require that at least some portion of IGMFs to arise from primordial sources predating the formation of stars. The presence of ${\cal B}_{\rm PMF}\simeq0.1{\rm\ nG}$ according to Ref.~\cite{Jedamzik:2020krr} could be sufficient to explain the Hubble tension. 

In this work our focus is to establish the Gilbertian (non-Amp\'erian = non-current) magnetic properties of the very dense $e^{+}e^{-}$ cosmic matter-antimatter plasma. In this framework, the magnetization of the early universe requires a large density of strong magnetic dipoles. Due to their large magnetic moment ($\propto e/m_e$) electrons and positrons magnetically dominate the universe. The dense $e^{+}e^{-}$-plasma is characterized in~\rf{fig:densityratio}: We show the antimatter (positron) abundance as a ratio to the prevailing baryon density as a function of cosmic photon temperature $T$. In this work we measure $T$ in units of energy (keV) thus we set the Boltzmann constant to $k_{B}=1$. We consider all results in temporal sequence in the expanding universe, thus we begin with high $T$ and early times on the left in~\rf{fig:densityratio} and end at lower $T$ and later times on the right.

\begin{figure}[ht]
 \centering
\includegraphics[width=0.45\textwidth]{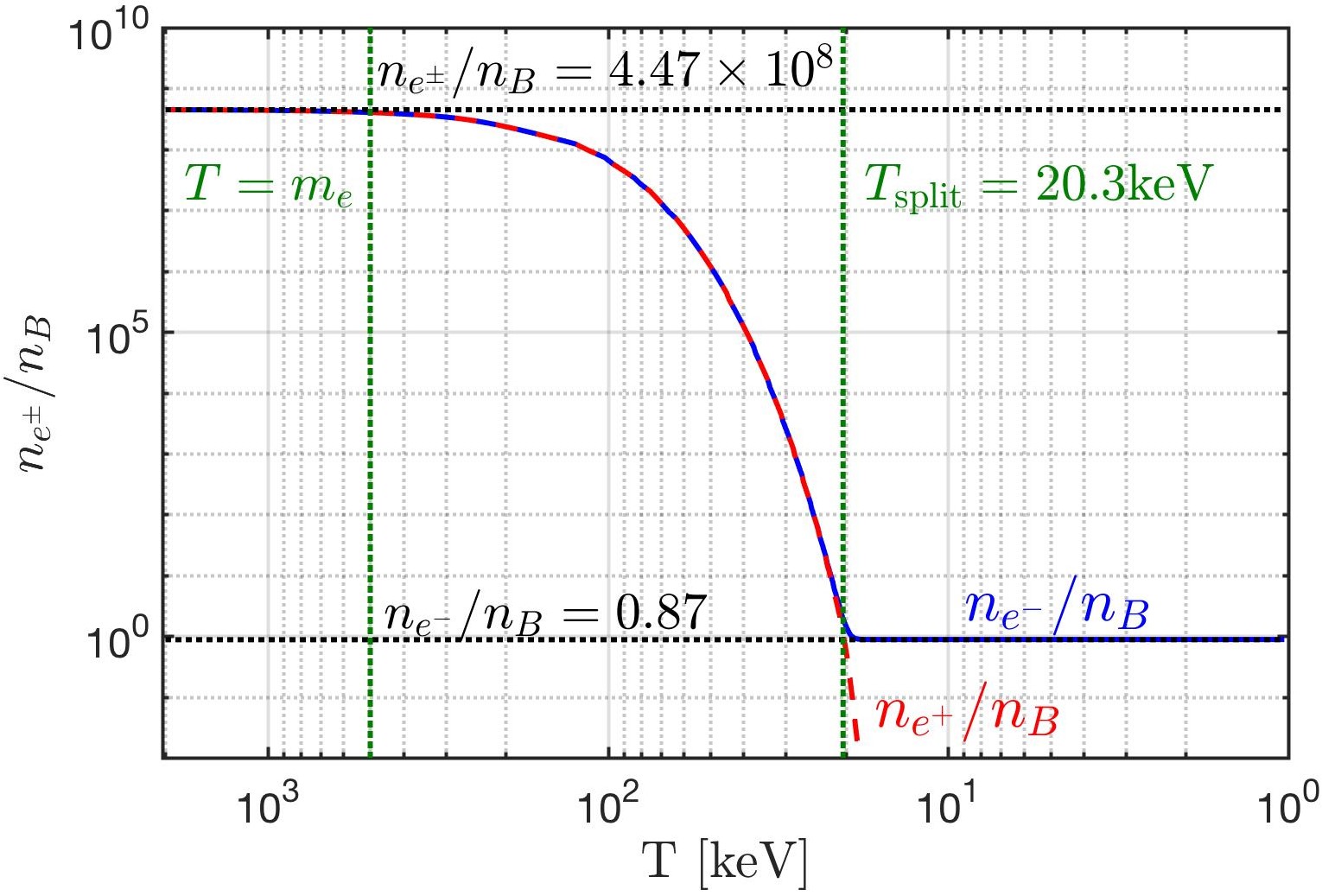}
 \caption{Number density of electron $e^{-}$ and positron $e^{+}$ to baryon ratio $n_{e^{\pm}}/n_{B}$ as a function of photon temperature in the universe. See text in~\rsec{sec:abundance} for further details.}
 \label{fig:densityratio} 
\end{figure}

We evaluate the magnetic moment polarization required for PMF magnitude of the spontaneous Gilbertian magnetization. Magnetic flux persistence implies that once the $e^{+}e^{-}$-pair plasma fades out, the ambient large scale Gilbertian magnetic field is maintained by the induced Amp\'erian (current) sources arising in the residual $e^{-}p^{+}\alpha^{++}$-plasma ultimately leading to the observed large scale structure IGMF. 

As we see in~\rf{fig:densityratio} at $T>m_ec^2=511\keV$ the $e^{+}e^{-}$-pair abundance was nearly 450 million pairs per baryon, dropping to about 100 million pairs per baryon at the pre-BBN temperature of $T=100\keV$. The number of $e^{+}e^{-}$-pairs is large compared to the residual `unpaired' electrons neutralizing the baryon charge locally down to $T_\mathrm{split}=20.3\keV$. Since electrons and positrons have opposite magnetic moments, the magnetized dense $e^{+}e^{-}$-plasma entails negligible net local spin density in statistical average. The residual very small polarization of unpaired electrons complements the magnetic field induced polarization of the proton component. 

As shown in Fig.\,2 in Ref.~\cite{Rafelski:2023emw}, following hadronization of the quark-gluon plasma (QGP) and below about $T\!=\!100\,000\keV$, in terms of energy densitythe early universe's first hour consists of photons, neutrinos and the $e^{+}e^{-}$-pair plasma. Massive dark matter and dark energy are negligible during this era. While we study the magnetic moment polarization of $e^{+}e^{-}$-plasma we do not address here its origin. However, we recall that the pair plasma decouples from the neutrino background near to $T=2000\keV$~\cite{Birrell:2014uka}. Therefore we consider the magnetic properties of the $e^{+}e^{-}$-pair plasma in the temperature range $2000\keV>T>20\keV$ and focus on the range $200\keV>T>20\keV$ where the most rapid antimatter abundance changes occurs and where the Boltzmann approximation is valid. This is notably the final epoch where antimatter exists in large quantities in the cosmos~\cite{Rafelski:2023emw}. 

The abundance of antimatter shown in~\rf{fig:densityratio} is obtained and discussed in more detail in~\rsec{sec:abundance}. Our analysis in~\rsec{sec:thermal} the four relativistic fermion gases (particle and antiparticle and both polarizations) where the spin and spin-orbit contributions are evaluated in~\rsec{sec:paradia}. The influence of magnetization on the charge chemical potential is determined in~\rsec{sec:chem}.  We show in~\rsec{sec:magnetization}, accounting for the matter-antimatter asymmetry present in the universe, that magnetization is nonzero. Our description of relativistic paramagnetism is covered in~\rsec{sec:paramagnetism}. The balance between paramagnetic and diamagnetic response is evaluated as a function of particle gyromagnetic ratio in~\rsec{sec:gfac}. The per-lepton magnetization is examined in~\rsec{sec:perlepton} distinguishing between cosmic and laboratory cases, in the latter case the number of magnetic dipoles is fixed, while in the universe the (comoving) number can  vary  with $T$. 

\rsec{sec:ferro} covers the consequences of forced magnetization via a magnetic moment polarization chemical potential. We find in~\rsec{sec:spinpot} that magnetization can be spontaneously increased in strength near the IGMF upper limit seen in~\req{igmf} given sufficient magnetic moment polarization. A model of self-magnetization is explored in~\rsec{sec:self} which indicates the need for flux conserving currents at low temperatures. Our findings are summarized in~\rsec{sec:conclusions}. We also suggest and a wealth of future follow-up projects mostly depending on introduction of transport theory that accounts for spin of particles in presence of a magnetic field.

\section{Cosmic electron-positron plasma abundance}
\label{sec:abundance}
\noindent As the universe cooled below temperature $T=m_{e}$ (the electron mass), the thermal electron and positron comoving density depleted by over eight orders of magnitude. At $T_{\rm split}=20.3\keV$, the charged lepton asymmetry (mirrored by baryon asymmetry and enforced by charge neutrality) became evident as the surviving excess electrons persisted while positrons vanished entirely from the particle inventory of the universe due to annihilation.

The electron-to-baryon density ratio $n_{e^{-}}/n_{B}$ is shown in~\rf{fig:densityratio} as the solid blue line while the positron-to-baryon ratio $n_{e^{+}}/n_{B}$ is represented by the dashed red line. These two lines overlap until the temperature drops below $T_{\rm split}=20.3\keV$ as positrons vanish from the universe marking the end of the $e^{+}e^{-}$-plasma and the dominance of the electron-proton $(e^{-}p)$-plasma. The two vertical dashed green lines denote temperatures $T=m_{e}\simeq511\keV$ and $T_{\rm split}=20.3\keV$. These results were obtained using charge neutrality and the baryon-to-photon content (entropy) of the universe; see details in~\cite{Rafelski:2023emw}. The two horizontal black dashed lines denote the relativistic $T\gg m_e$ abundance of $n_{e^{\pm}}/n_{B}=4.47\times10^{8}$ and post-annihilation abundance of $n_{e^{-}}/n_{B}=0.87$. Above temperature $T\simeq85\keV$, the $e^{+}e^{-}$ primordial plasma density exceeded that of the Sun's core density $n_{e}\simeq6\times10^{26}{\rm\ cm}^{-3}$~\cite{Bahcall:2000nu}. 

Conversion of the dense $e^{+}e^{-}$-pair plasma into photons reheated the photon background~\cite{Birrell:2014uka} separating the photon and neutrino temperatures. The $e^{+}e^{-}$ annihilation and photon reheating period lasted no longer than an afternoon lunch break. Because of charge neutrality, the post-annihilation comoving ratio $n_{e^{-}}/n_{B}=0.87$~\cite{Rafelski:2023emw} is slightly offset from unity in~\rf{fig:densityratio} by the presence of bound neutrons in $\alpha$ particles and other neutron containing light elements produced during BBN epoch. 

To obtain a quantitative description of the above evolution, we study the bulk properties of the relativistic charged/magnetic gasses in a nearly homogeneous and isotropic primordial universe via the thermal Fermi-Dirac or Bose distributions. For matter $(e^{-};\ \sigma=+1)$ and antimatter $(e^{+};\ \sigma=-1)$ particles, a nonzero relativistic chemical potential $\mu_{\sigma}=\sigma\mu$ is caused by an imbalance of matter and antimatter. While the primordial electron-positron plasma era was overall charge neutral, there was a small asymmetry in the charged leptons from baryon asymmetry~\cite{Fromerth:2012fe,Canetti:2012zc} in the universe. Reactions such as $e^{+}e^{-}\leftrightarrow\gamma\gamma$ constrains the chemical potential of electrons and positrons~\cite{Elze:1980er} as 
\begin{align}
 \label{cpotential}
 \mu\equiv\mu_{e^{-}}=-\mu_{e^{+}}\,,\qquad
 \lambda\equiv\lambda_{e^{-}}=\lambda_{e^{+}}^{-1}=\exp\frac{\mu}{T}\,,
\end{align}
where $\lambda$ is the fugacity of the system. 

During the $e^{+}e^{-}$-plasma epoch, the density changed dramatically over time (see~\rf{fig:densityratio}) changing the chemical potential in turn. We can then parameterize the chemical potential of the $e^{+}e^{-}$-plasma as a function of temperature $\mu\rightarrow\mu(T)$ via the charge neutrality of the universe which implies
\begin{align}
 \label{chargeneutrality}
 n_{p}=n_{e^{-}}-n_{e^{+}}=\frac{1}{V}\lambda\frac{\partial}{\partial\lambda}\ln{\cal Z}_{e^{+}e^{-}}\,.
\end{align}
In~\req{chargeneutrality}, $n_{p}$ is the observed total number density of protons in all baryon species. The parameter $V$ relays the proper volume under consideration and $\ln{\cal Z}_{e^{+}e^{-}}$ is the partition function for the electron-positron gas. The chemical potential defined in~\req{cpotential} is obtained from the requirement that the positive charge of baryons (protons, $\alpha$ particles, light nuclei produced after BBN) is exactly and locally compensated by a tiny net excess of electrons over positrons.

The abundance of baryons is itself fixed by the known abundance relative to photons~\cite{ParticleDataGroup:2022pth} and we employed the contemporary recommended value $n_B/n_\gamma=6.09\times 10^{-10}$. The resulting chemical potential needs to be evaluated carefully to obtain the behavior near to $T_{\rm split}=20.3\keV$ where the relatively small value of chemical potential $\mu$ rises rapidly so that positrons vanish from the particle inventory of the universe while nearly one electron per baryon remains. The detailed solution of this problem is found in Refs.\;\cite{Fromerth:2012fe,Rafelski:2023emw} leading to the results shown in~\rf{fig:densityratio}. These results are obtained allowing for Fermi-Dirac and Bose statistics, however it is often numerically sufficient to consider the Boltzmann distribution limit; see~\rsec{sec:paradia}.

The partition function of the $e^{+}e^{-}$-plasma can be understood as the sum of four gaseous species
\begin{align}
 \label{partition:0} 
 \ln{\cal Z}_{e^{+}e^{-}}=\ln{\cal Z}_{e^{+}}^{\uparrow}+\ln{\cal Z}_{e^{+}}^{\downarrow}+\ln{\cal Z}_{e^{-}}^{\uparrow}+\ln{\cal Z}_{e^{-}}^{\downarrow}\,,
\end{align}
of electrons and positrons of both polarizations. In the presence of a magnetic field ${\cal B}$ along a primary axis, there is some modification of the usual relativistic fermion partition function which is now given by
\begin{align}
 \begin{split}
 \label{partition:1}
 \ln{\cal Z}_{e^{+}e^{-}}&=\frac{e{\cal B}V}{(2\pi)^{2}}\sum_{\sigma}^{\pm1}\sum_{s}^{\pm1}\sum_{n=0}^{\infty}\int_{-\infty}^{\infty}{\rm d}p_{z}\\
 &\left[\ln\left(1+\lambda_{\sigma}\xi_{\sigma,s}\exp\left(-\frac{E}{T}\right)\right)\right]\, 
 \end{split}\\
 \label{partition:2} 
 \lambda_{\sigma}\xi_{\sigma, s} &= \exp{\frac{\mu_{\sigma}+\eta_{\sigma,s}}{T}}\,,
\end{align}
where $p_{z}$ is the momentum parallel to the field axis and electric charge is $e\equiv q_{e^{+}}=-q_{e^{-}}$. The index $\sigma$ in~\req{partition:1} is a sum over electron and positron states while $s$ is a sum over polarizations. The index $s$ refers to the spin along the field axis: parallel $(\uparrow;\ s=+1)$ or anti-parallel $(\downarrow;\ s=-1)$ for both particle and antiparticle species.

As the gas is electrically neutral, we will for the time being ignore charge-charge interactions. There is an additional deformation of the distribution from particle creation and destruction correlations; see Ch.~11 of~\cite{Letessier:2002ony} in the context of quark flavors. These will be not included as the considering volume is always large. The quantum numbers of the energy eigenstate $E$ will be elaborated on in~\rsec{sec:thermal}.

We are explicitly interested in small asymmetries such as baryon excess over antibaryons, or one polarization over another. These are described by~\req{partition:2} as the following two fugacities:
\begin{itemize}
 \item[a.] Chemical fugacity $\lambda_{\sigma}$
 \item[b.] Polarization fugacity $\xi_{\sigma,s}$
\end{itemize}
The chemical fugacity $\lambda_{\sigma}$ (defined in~\req{cpotential} above) describes deformation of the Fermi-Dirac distribution due to nonzero chemical potential $\mu$. An imbalance in electrons and positrons leads as discussed earlier to a nonzero particle chemical potential $\mu\neq0$. We then introduce a novel polarization fugacity $\xi_{\sigma,s}$ and polarization potential $\eta_{\sigma,s}=\sigma s\eta$. We propose the polarization potential follows analogous expressions as seen in~\req{cpotential} obeying
\begin{align}
 \label{spotential}
 \eta\equiv\eta_{+,+}=\eta_{-,-}\,,\quad\eta=-\eta_{\pm,\mp}\,,\quad\xi_{\sigma,s}\equiv\exp{\frac{\eta_{\sigma,s}}{T}}\,.
\end{align}

An imbalance in polarization within a region of volume $V$ results in a nonzero magnetic moment potential $\eta\neq0$. Conveniently since antiparticles have opposite sign of charge and magnetic moment, the same magnetic moment is associated with opposite spin orientation for particles and antiparticles independent of degree of spin-magnetization. A completely particle-antiparticle symmetric magnetized plasma will have therefore zero total angular momentum. This is of course very different from the situation today of a matter dominated universe.

\section{Theory of magnetized matter-antimatter plasmas}
\label{sec:thermal}
\noindent As the universe undergoes isotropic expansion, the temperature decreases adiabatically~\cite{Abdalla:2022yfr} and conserves entropy as 
\begin{align}
 \label{tscale}
 T(t)=T_{0}\frac{a_{0}}{a(t)}\rightarrow T(z)=T_{0}(1+z)\,,
\end{align}
where $a(t)$ is the scale factor defined by the Friedmann-Lema{\^i}tre-Robertson-Walker (FLRW) metric~\cite{weinberg1972gravitation} and $z$ is the redshift. The comoving temperature $T_{0}$ is given by the present day temperature of the CMB, with contemporary scale factor $a_{0}=1$. Within a homogeneous magnetic domain, the magnetic field magnitude varies~\cite{Durrer:2013pga} over cosmic expansion as
\begin{align}
 \label{bscale}
 {\cal B}(t)={\cal B}_{0}\frac{a_{0}^{2}}{a^{2}(t)}\rightarrow{\cal B}(z)={\cal B}_{0}\left(1+z\right)^{2}\,,
\end{align}
where ${\cal B}_{0}$ is the comoving value of the magnetic field obtained from the contemporary value today given in~\req{igmf}. Non-primordial magnetic fields (which are generated through other mechanisms such as dynamo or astrophysical sources) do not follow this scaling~\cite{Pomakov:2022cem}. The presence of matter and late universe structure formation also contaminates the primordial field evolution in~\req{bscale}. It is only in deep intergalactic space where primordial fields remain preserved and comoving over cosmic time.

From~\req{tscale} and~\req{bscale} emerges a natural ratio of interest here which is conserved over cosmic expansion 
\begin{gather}
 \label{tbscale}
 b \equiv\frac{e{\cal B}(t)}{T^{2}(t)}=\frac{e{\cal B}_{0}}{T_{0}^{2}}\equiv b_0={\rm\ const.}\\
 10^{-3}>b_{0}>10^{-11}\,,
\end{gather}
given in natural units ($c=\hbar=k_{B}=1$). We computed the bounds for this cosmic magnetic scale ratio by using the present day IGMF observations given by~\req{igmf} and the present CMB temperature $T_{0}=2.7{\rm\ K}\simeq2.3\times10^{-4}\eV$~\cite{Planck:2018vyg}.

To evaluate magnetic properties of the thermal $e^{+}e^{-}$-pair plasma we take inspiration from Ch. 9 of Melrose's treatise on magnetized plasmas~\cite{melrose2008quantum}. We focus on the bulk properties of thermalized plasmas in (near) equilibrium. In considering $e^{+}e^{-}$-pair plasma, we introduce the microscopic energy of the charged relativistic fermion within a homogeneous ($z$-direction) magnetic field~\cite{Steinmetz:2018ryf}. The energy eigenvalue is given by
\begin{align}
 \label{kgp}
 E^{n}_{\sigma,s}(p_{z},{\cal B})=\sqrt{m_{e}^{2}+p_{z}^{2}+e{\cal B}\left(2n+1+\frac{g}{2}\sigma s\right)}\,,
\end{align}
where $n\in0,1,2,\ldots$ is the Landau orbital quantum number.~\req{kgp} differentiates between electrons and positrons which is to ensure the correct non-relativistic limit is reached; see~\rf{fig:schematic}. The parameter $g$ is the gyro-magnetic ($g$-factor) of the particle. Following the conventions found in~\cite{Tiesinga:2021myr}, we set $g\equiv g_{e^{+}}=-g_{e^{-}}>0$ such that electrons and positrons have opposite $g$-factors and opposite magnetic moments which is schematically shown in~\rf{fig:schematic}.

As statistical properties depend on the characteristic Boltzmann factor $E/T$, another interpretation of~\req{tbscale} in the context of energy eigenvalues (such as those given in~\req{kgp}) is the preservation of magnetic moment energy relative to momentum under adiabatic cosmic expansion.

We rearrange~\req{kgp} by pulling the spin dependency and the ground state Landau orbital into the mass writing
\begin{gather}
 \label{effmass:1}
 E^{n}_{\sigma,s}={\tilde m}_{\sigma,s}\sqrt{1+\frac{p_{z}^{2}}{{\tilde m}_{\sigma,s}^{2}}+\frac{2e{\cal B}n}{{\tilde m}_{\sigma,s}^{2}}}\,,\\
 \label{effmass:2}
 \varepsilon_{\sigma,s}^{n}(p_{z},{\cal B})=\frac{E_{\sigma,s}^{n}}{{\tilde m}_{\sigma,s}}\,,\qquad{\tilde m}_{\sigma,s}^{2}=m_{e}^{2}+e{\cal B}\left(1+\frac{g}{2}\sigma s\right)\,,
\end{gather}
where we introduced the dimensionless energy $\varepsilon^{n}_{\sigma,s}$ and effective polarized mass ${\tilde m}_{\sigma,s}$ which is distinct for each spin alignment and is a function of magnetic field strength ${\cal B}$. The effective polarized mass ${\tilde m}_{\sigma,s}$ allows us to describe the $e^{+}e^{-}$-plasma with the spin effects almost wholly separated from the Landau characteristics of the gas when considering the plasma's thermodynamic properties.

\begin{figure}[ht]
 \centering
 \includegraphics[width=0.4\textwidth]{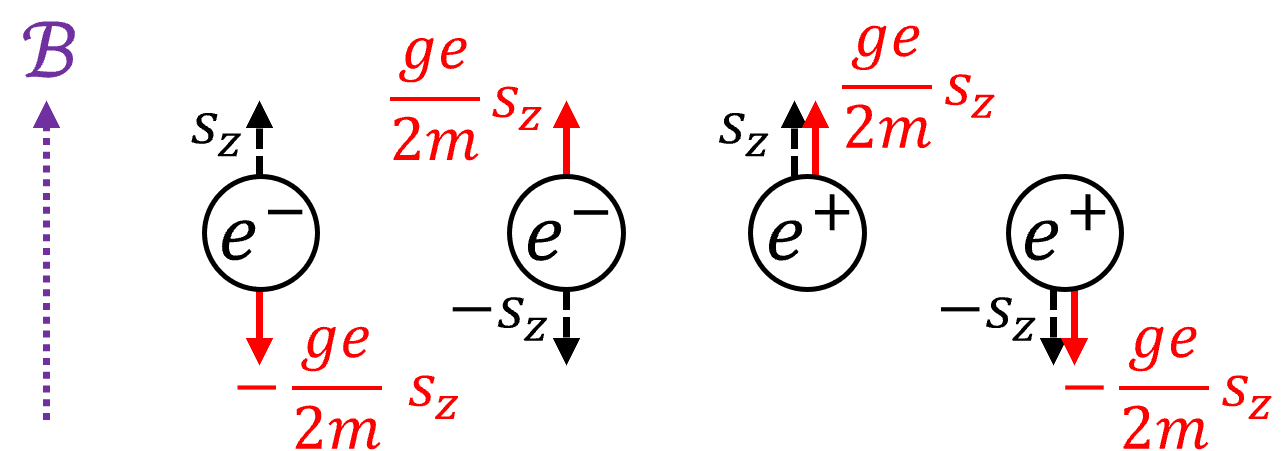}\Bstrut\\
 \begin{tabular}{ r|c|c| }
 \multicolumn{1}{r}{}
 & \multicolumn{1}{c}{aligned: $s=+1$}
 & \multicolumn{1}{c}{anti-aligned: $s=-1$} \\
 \cline{2-3}
 electron: $\sigma=+1$ & $U_{\rm Mag}>0$ & $U_{\rm Mag}<0$ \TBstrut\\
 \cline{2-3}
 positron: $\sigma=-1$ & $U_{\rm Mag}<0$ & $U_{\rm Mag}>0$ \TBstrut\\
 \cline{2-3}
 \end{tabular}\\\,\TBstrut\\
 \begin{tabular}{ r|c|c| }
 \multicolumn{1}{r}{}
 & \multicolumn{1}{c}{aligned: $s=+1$}
 & \multicolumn{1}{c}{anti-aligned: $s=-1$} \\
 \cline{2-3}
 electron: $\sigma=+1$ & $+\mu+\eta$ & $+\mu-\eta$ \TBstrut\\
 \cline{2-3}
 positron: $\sigma=-1$ & $-\mu-\eta$ & $-\mu+\eta$ \TBstrut\\
 \cline{2-3}
 \end{tabular}
 \caption{Organizational schematic of matter-antimatter $(\sigma)$ and polarization $(s)$ states with respect to the sign of the non-relativistic magnetic dipole energy $U_{\rm Mag}$ (obtainable from~\req{kgp}) and the chemical $\mu$ and polarization $\eta$ potentials as seen in~\req{partitionpower:2}.}
 \label{fig:schematic}
\end{figure}


Since we address the temperature interval $200\keV>T>20\keV$ where the effects of quantum Fermi statistics on the $e^{+}e^{-}$-pair plasma are relatively small, but the gas is still considered relativistic, we will employ the Boltzmann approximation to the partition function in~\req{partition:1}. However, we extrapolate our results for presentation completeness up to $T\simeq 4m_{e}$.

In general, modifications due to quantum statistical phase-space reduction for fermions are expected to suppress results by about 20\% in the extrapolated regions. We will continue to search for semi-analytical solutions for Fermi statistics in relativistic $e^{+}e^{-}$-pair gasses to compliment the Boltzmann solution offered here. 

\subsection{Unified treatment of para and diamagnetism}
\label{sec:paradia}
\noindent We will proceed in this section with the Boltzmann approximation for the limit where $T\lesssim m_e$. The partition function shown in equation~\req{partition:1} can be rewritten removing the logarithm as
\begin{align}
\begin{split}
\label{partitionpower:1}
\ln{{\cal Z}_{e^{+}e^{-}}}&=\frac{e{\cal B}V}{(2\pi)^{2}}\sum_{\sigma,s}^{\pm1}\sum_{n=0}^{\infty}\sum_{k=1}^{\infty}\int_{-\infty}^{+\infty}{\rm d}p_{z}\\
&\frac{(-1)^{k+1}}{k}\exp\left({k\frac{\sigma\mu+\sigma s\eta-{\tilde m}_{\sigma,s}\varepsilon^{n}_{\sigma,s}}{T}}\right)\,, 
\end{split}\\
\label{bapprox} 
&\sigma\mu+\sigma s\eta-{\tilde m}_{\sigma,s}\varepsilon_{\sigma,s}^{n}<0\,,
\end{align}
which is well behaved as long as the factor in~\req{bapprox} remains negative. We evaluate the sums over $\sigma$ and $s$ as
\begin{multline}
 \label{partitionpower:2}
 \ln{{\cal Z}_{e^{+}e^{-}}}=\frac{e{\cal B}V}{(2\pi)^{2}}\sum_{n=0}^{\infty}\sum_{k=1}^{\infty}\int_{-\infty}^{+\infty}{\rm d}p_{z}\frac{(-1)^{k+1}}{k}\\
 \left(\ \exp\left(k\frac{+\mu+\eta}{T}\right)\exp\left(-k\frac{{\tilde m}_{+,+}\varepsilon_{+,+}^{n}}{T}\right)\right.\\
 +\exp\left(k\frac{+\mu-\eta}{T}\right)\exp\left(-k\frac{{\tilde m}_{+,-}\varepsilon_{+,-}^{n}}{T}\right)\\
 +\exp\left(k\frac{-\mu-\eta}{T}\right)\exp\left(-k\frac{{\tilde m}_{-,+}\varepsilon_{-,+}^{n}}{T}\right)\\
 +\left.\exp\left(k\frac{-\mu+\eta}{T}\right)\exp\left(-k\frac{{\tilde m}_{-,-}\varepsilon_{-,-}^{n}}{T}\right)\right)\,.
\end{multline}
We note from~\rf{fig:schematic} that the first and forth terms and the second and third terms share the same energies via
\begin{gather}
 \label{partitionpower:3}
 \varepsilon_{+,+}^{n}=\varepsilon_{-,-}^{n}\,,\qquad
 \varepsilon_{+,-}^{n}=\varepsilon_{-,+}^{n}\,.\qquad
 \varepsilon_{+,-}^{n}<\varepsilon_{+,+}^{n}\,.
\end{gather}

\req{partitionpower:3} allows us to reorganize the partition function with a new magnetization quantum number $s'$ which characterizes paramagnetic flux increasing states $(s'=+1)$ and diamagnetic flux decreasing states $(s'=-1)$. This recasts~\req{partitionpower:2} as
\begin{multline}
 \label{partitionpower:4}
 \ln{{\cal Z}_{e^{+}e^{-}}}=\frac{e{\cal B}V}{(2\pi)^{2}}\sum_{s'}^{\pm1}\sum_{n=0}^{\infty}\sum_{k=1}^{\infty}\int_{-\infty}^{+\infty}{\rm d}p_{z}\frac{(-1)^{k+1}}{k}\\
 \left[2\xi_{s'}\cosh\frac{k\mu}{T}\right]\exp\left(-k\frac{{\tilde m}_{s'}\varepsilon_{s'}^{n}}{T}\right)\,,
\end{multline}
with dimensionless energy, polarization mass, and polarization redefined in terms of $s'$
\begin{gather}
 \epsilon_{s'=+1}^{n}=\epsilon_{+,-}^{n}\,,\qquad
 \epsilon_{s'=-1}^{n}=\epsilon_{+,+}^{n}\,,\\
 {\tilde m}_{s'}^{2}=m_{e}^{2}+e{\cal B}\left(1-\frac{g}{2}s'\right)\,,\\
 \eta\equiv\eta_{+}=-\eta_{-}\qquad\xi\equiv\xi_{+}=\xi_{-}^{-1}\,.
\end{gather}

We introduce the modified Bessel function $K_{\nu}$ (see Ch. 10 of~\cite{Letessier:2002ony}) of the second kind
\begin{align}
\begin{split}
\label{besselk}
K_{\nu}\left(\frac{m}{T}\right)&=\frac{\sqrt{\pi}}{\Gamma(\nu-1/2)}\frac{1}{m}\left(\frac{1}{2mT}\right)^{\nu-1}\\
&\int_{0}^{\infty}{\rm d}p\,p^{2\nu-2}\exp\left({-\frac{m\varepsilon}{T}}\right)\,,
\end{split}\\
\nu&>1/2\,,\qquad\varepsilon=\sqrt{1+p^{2}/m^{2}}\,,
\end{align}
allowing us to rewrite the integral over momentum in~\req{partitionpower:4} as
\begin{align}
 \label{besselkint}
 \frac{1}{T}\int_{0}^{\infty}\!\!{\rm d}p_{z}\exp\!\left(\!{-\frac{k{\tilde m}_{s'}\varepsilon_{s'}^{n}}{T}}\!\right)\!=\!W_{1}\!\!\left(\frac{k{\tilde m}_{s'}\varepsilon_{s'}^{n}(0,{\cal B})}{T}\right)\,.
\end{align}
The function $W_{\nu}$ serves as an auxiliary function of the form $W_{\nu}(x)=xK_{\nu}(x)$. The notation $\varepsilon(0,{\cal B})$ in~\req{besselkint} refers to the definition of dimensionless energy found in~\req{effmass:2} with $p_{z}=0$.

The standard Boltzmann distribution is obtained by summing only $k=1$ and neglecting the higher order terms. The Euler-Maclaurin formula~\cite{abramowitz1988handbook} is used to convert the summation over Landau levels into an integration given by
\begin{multline}
 \label{eulermaclaurin}
 \sum_{n=0}^{\infty}W_{1}(n)=\int_{0}^{\infty}{\rm d}n\,W_{1}(n)+\frac{1}{2}\left[W_{1}(\infty)+W_{1}(0)\right]\\
 +\frac{1}{12}\left[\frac{\partial W_{1}}{\partial n}\bigg\vert_{\infty}-\frac{\partial W_{1}}{\partial n}\bigg\vert_{0}\right]+{\cal R}
\end{multline}
where ${\cal R}$ is the resulting power series and error remainder of the integration defined in terms of Bernoulli polynomials. Euler-Maclaurin integration is rarely convergent, and in this case serves only as an approximation within the domain where the error remainder is small and bounded; see Ref.~\cite{greiner2012thermodynamics} for the non-relativistic case. In this analysis, we keep the zeroth and first order terms in the Euler-Maclaurin formula. We note that regularization of the excess terms in~\req{eulermaclaurin} is done in the context of strong field QED~\cite{greiner2008quantum} though that is outside our scope.

After truncation of the series and error remainder and combining~\req{partitionpower:1} through~\req{eulermaclaurin}, the partition function can then be written in terms of modified Bessel $K_{\nu}$ functions of the second kind, yielding
\begin{align}
 \begin{split}
 \label{boltzmann}
 \ln{\cal Z}_{e^{+}e^{-}}&\simeq\frac{T^{3}V}{\pi^{2}}\sum_{s'}^{\pm1}\left[\xi_{s'}\cosh{\frac{\mu}{T}}\right]\\
 &\left(x_{s'}^{2}K_{2}(x_{s'})+\frac{b_{0}}{2}x_{s'}K_{1}(x_{s'})+\frac{b_{0}^{2}}{12}K_{0}(x_{s'})\right)\,,
 \end{split}\\
 \label{xfunc}
 x_{s'}&=\frac{{\tilde m}_{s'}}{T}=\sqrt{\frac{m_{e}^{2}}{T^{2}}+b_{0}\left(1-\frac{g}{2}s'\right)}\,.
\end{align}
The latter two terms in~\req{boltzmann} proportional to $b_{0}K_{1}$ and $b_{0}^{2}K_{0}$ are the uniquely magnetic terms present containing both spin and Landau orbital influences in the partition function. The $K_{2}$ term is analogous to the textbook-case of free Fermi gas~\cite{greiner2012thermodynamics}, being modified only by spin effects.

This \lq separation of concerns\rq\ can be rewritten as
\begin{align}
 \label{spin}
 \ln{\cal Z}_{\rm S}&=\frac{T^{3}V}{\pi^{2}}\sum_{s'}^{\pm1}\left[\xi_{s'}\cosh{\frac{\mu}{T}}\right]\left(x_{s'}^{2}K_{2}(x_{s'})\right)\,,\\
 \begin{split}
 \label{spinorbit}
 \ln{\cal Z}_{\rm SO}&=\frac{T^{3}V}{\pi^{2}}\sum_{s'}^{\pm}\left[\xi_{s'}\cosh{\frac{\mu}{T}}\right]\\
 &\left(\frac{b_{0}}{2}x_{s'}K_{1}(x_{s'})+\frac{b_{0}^{2}}{12}K_{0}(x_{s'})\right)\,, 
 \end{split}
\end{align}

where the spin (S) and spin-orbit (SO) partition functions can be considered independently. When the magnetic scale $b_{0}$ is small, the spin-orbit term~\req{spinorbit} becomes negligible leaving only paramagnetic effects in~\req{spin} due to spin. In the non-relativistic limit,~\req{spin} reproduces a quantum gas whose Hamiltonian is defined as the free particle (FP) Hamiltonian plus the magnetic dipole (MD) Hamiltonian which span two independent Hilbert spaces ${\cal H}_{\rm FP}\otimes{\cal H}_{\rm MD}$.

Writing the partition function as~\req{boltzmann} instead of~\req{partitionpower:1} has the additional benefit that the partition function remains finite in the free gas $({\cal B}\rightarrow0)$ limit. This is because the free Fermi gas and~\req{spin} are mathematically analogous to one another. As the Bessel $K_{\nu}$ functions are evaluated as functions of $x_{\pm}$ in~\req{xfunc}, the \lq free\rq\ part of the partition $K_{2}$ is still subject to dipole magnetization effects. In the limit where ${\cal B}\rightarrow0$, the free Fermi gas is recovered in both the Boltzmann approximation $k=1$ and the general case $\sum_{k=1}^{\infty}$.

\subsection{Charge chemical potential response}
\label{sec:chem}
\noindent In presence of a magnetic field in the Boltzmann approximation, the charge neutrality condition~\req{chargeneutrality} becomes
\begin{multline}
 \label{chem}
 \sinh\frac{\mu}{T}=n_{p}\frac{\pi^{2}}{T^{3}}\\
 \left[\sum_{s'}^{\pm1}\xi_{s'}\!\left(\!x_{s'}^{2}K_{2}(x_{s'})\!+\!\frac{b_{0}}{2}x_{s'}K_{1}(x_{s'})\!+\!\frac{b_{0}^{2}}{12}K_{0}(x_{s'}\!)\!\right)\!\right]^{-1}\!.
\end{multline}
\req{chem} is fully determined by the right-hand-side expression if the magnetic moment fugacity is set to unity $\eta=0$ implying no external bias to the number of polarizations except as a consequence of the difference in energy eigenvalues. In practice, the latter two terms in~\req{chem} are negligible to chemical potential in the bounds of the primordial $e^{+}e^{-}$-plasma considered and only becomes relevant for extreme (see~\rf{fig:chemicalpotential}) magnetic field strengths well outside our scope.

\req{chem} simplifies if there is no external magnetic field $b_{0}=0$ into
\begin{align}
 \label{simpchem:1}
 \sinh\frac{\mu}{T}=n_{p}\frac{\pi^{2}}{T^{3}}\left[2\cosh\frac{\eta}{T}\left(\frac{m_{e}}{T}\right)^{2}K_{2}\left(\frac{m_{e}}{T}\right)\right]^{-1}\,.
\end{align}

\begin{figure}[ht]
 \centering
 \includegraphics[width=0.45\textwidth]{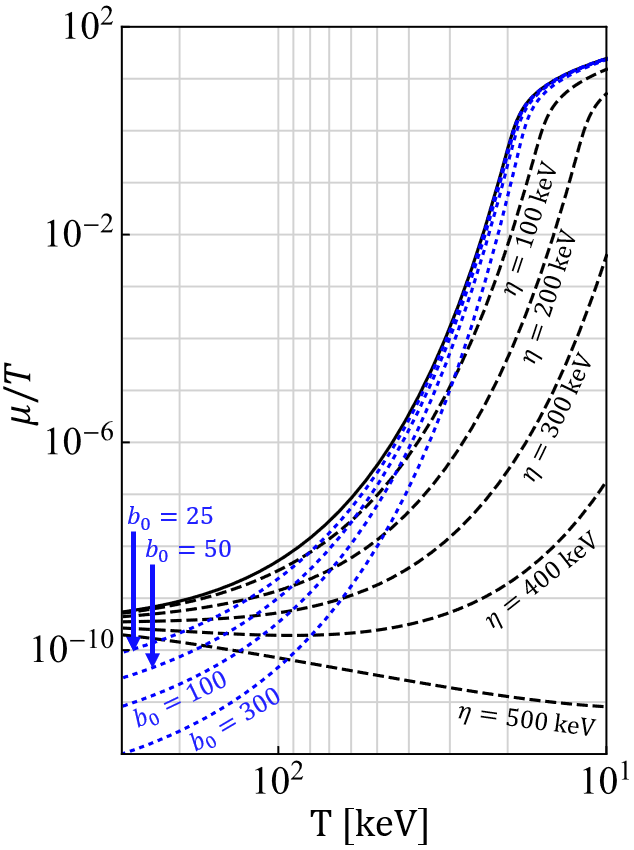}
 \caption{The chemical potential over temperature $\mu/T$ is plotted as a function of temperature with differing values of magnetic moment potential $\eta$ and magnetic scale $b_{0}$.}
 \label{fig:chemicalpotential}
\end{figure}

In~\rf{fig:chemicalpotential} we plot the chemical potential $\mu/T$ in~\req{chem} and~\req{simpchem:1} which characterizes the importance of the charged lepton asymmetry as a function of temperature. Since the baryon (and thus charged lepton) asymmetry remains fixed, the suppression of $\mu/T$ at high temperatures indicates a large pair density which is seen explicitly in~\rf{fig:densityratio}. The black line corresponds to the $b_{0}=0$ and $\eta=0$ case. 

The para-diamagnetic contribution from~\req{spinorbit} does not appreciably influence $\mu/T$ until the magnetic scales involved become incredibly large well outside the observational bounds defined in~\req{igmf} and~\req{tbscale} as seen by the dotted blue curves of various large values $b_{0}=\{25,\ 50,\ 100,\ 300\}$. The chemical potential is also insensitive to forcing by the magnetic moment potential until $\eta$ reaches a significant fraction of the electron mass $m_{e}$ in size. The chemical potential for large values of magnetic moment potential $\eta=\{100,\ 200,\ 300,\ 400,\ 500\}\,\keV$ are also plotted as dashed black lines with $b_{0}=0$.

It is interesting to note that there are crossing points where a given chemical potential can be described as either an imbalance in magnetic moment polarization or presence of external magnetic field. While magnetic moment potential suppresses the chemical potential at low temperatures, external magnetic fields only suppress the chemical potential at high temperatures.

The profound insensitivity of the chemical potential to these parameters justifies the use of the free particle chemical potential (black line) in the ranges of magnetic field strength considered for cosmology. Mathematically this can be understood as $\xi$ and $b_{0}$ act as small corrections in the denominator of~\req{chem} if expanded in powers of these two parameters.

\section{Gilbertian magnetization of electron-positron plasma}
\label{sec:magnetization}
\noindent The total magnetic flux within a region of space can be written as the sum of external fields and the magnetization of the medium via
\begin{align}
 \label{totalmag}
 {\cal B}_{\rm total} = {\cal B} + {\cal M}\,.
\end{align}
For the simplest mediums without ferromagnetic or hysteresis considerations, the relationship can be parameterized by the susceptibility $\chi$ of the medium as
\begin{align}
 \label{susceptibility}
 {\cal B}_{\rm total} = (1+\chi){\cal B}\,,\qquad {\cal M} = \chi{\cal B}\,,
\end{align}
with the possibility of both paramagnetic materials $(\chi>1)$ and diamagnetic materials $(\chi<1)$. The $e^{+}e^{-}$-plasma however does not so neatly fit in either category as given by~\req{spin} and~\req{spinorbit}. In general, the susceptibility of the gas will itself be a field dependant quantity given by
\begin{align}
 \chi \equiv \frac{\partial{\cal M}}{\partial{\cal B}}\,.
\end{align}

In our analysis, the external magnetic field always appears within the context of the magnetic scale $b_{0}$, therefore we can introduce the change of variables
\begin{align}
 \frac{\partial b_{0}}{\partial{\cal B}}=\frac{e}{T^{2}}\,.
\end{align}
The magnetization of the $e^{+}e^{-}$-plasma described by the partition function in~\req{boltzmann} can then be written as
\begin{align}
 \label{defmagetization}
 {\cal M}\equiv\frac{T}{V}\frac{\partial}{\partial{\cal B}}\ln{{\cal Z}_{e^{+}e^{-}}} = \frac{T}{V}\left(\frac{\partial b_{0}}{\partial{\cal B}}\right)\frac{\partial}{\partial b_{0}}\ln{{\cal Z}_{e^{+}e^{-}}}\,,
\end{align}
Magnetization arising from other components in the cosmic gas (protons, neutrinos, etc.) could in principle also be included. Localized inhomogeneities of matter evolution are often non-trivial and generally be solved numerically using magneto-hydrodynamics (MHD)~\cite{melrose2008quantum,Vazza:2017qge,Vachaspati:2020blt}. In the context of MHD, primordial magnetogenesis from fluid flows in the electron-positron epoch was considered in~\cite{Gopal:2004ut,Perrone:2021srr}.

We introduce dimensionless units for magnetization ${\mathfrak M}$ by defining the critical field strength
\begin{align}
 {\cal B}_{C}\equiv\frac{m_{e}^{2}}{e}\,,\qquad{\mathfrak M}\equiv\frac{\cal M}{{\cal B}_{C}}\,.
\end{align}
The scale ${\cal B}_{C}$ is where electromagnetism is expected to become subject to non-linear effects, though luckily in our regime of interest, electrodynamics should be linear. We note however that the upper bounds of IGMFs in~\req{igmf} (with $b_{0}=10^{-3}$; see~\req{tbscale}) brings us to within $1\%$ of that limit for the external field strength in the temperature range considered.

The total magnetization ${\mathfrak M}$ can be broken into the sum of magnetic moment parallel ${\mathfrak M}_{+}$ and magnetic moment anti-parallel ${\mathfrak M}_{-}$ contributions
\begin{align}
\label{g2mag}
{\mathfrak M}={\mathfrak M}_{+}+{\mathfrak M}_{-}\,.
\end{align}
We note that the expression for the magnetization simplifies significantly for $g=2$ which is the \lq natural\rq\ gyro-magnetic factor~\cite{Evans:2022fsu,Rafelski:2022bsv} for Dirac particles. For illustration, the $g=2$ magnetization from~\req{defmagetization} is then
\begin{align}
 \label{g2magplus}
 {\mathfrak M}_{+}&=\frac{e^{2}}{\pi^{2}}\frac{T^{2}}{m_{e}^{2}}\xi\cosh{\frac{\mu}{T}}\left[\frac{1}{2}x_{+}K_{1}(x_{+})+\frac{b_{0}}{6}K_{0}(x_{+})\right]\,,\\
\begin{split} 
 \label{g2magminus}
 -{\mathfrak M}_{-}&=\frac{e^{2}}{\pi^{2}}\frac{T^{2}}{m_{e}^{2}}\xi^{-1}\cosh{\frac{\mu}{T}}\\
 &\left[\left(\frac{1}{2}+\frac{b_{0}^{2}}{12x_{-}^{2}}\right)x_{-}K_{1}(x_{-})+\frac{b_{0}}{3}K_{0}(x_{-})\right]\,,
\end{split}\\
 x_{+}&=\frac{m_{e}}{T}\,,\qquad
 x_{-}=\sqrt{\frac{m_{e}^{2}}{T^{2}}+2b_{0}}\,.
\end{align}
As the $g$-factor of the electron is only slightly above two at $g\simeq2.00232$~\cite{Tiesinga:2021myr}, the above two expressions for ${\mathfrak M}_{+}$ and ${\mathfrak M}_{-}$ are only modified by a small amount because of anomalous magnetic moment (AMM) and would be otherwise invisible on our figures. We will revisit AMM in~\rsec{sec:gfac}.

\begin{figure}[ht]
 \centering
 \includegraphics[width=0.45\textwidth]{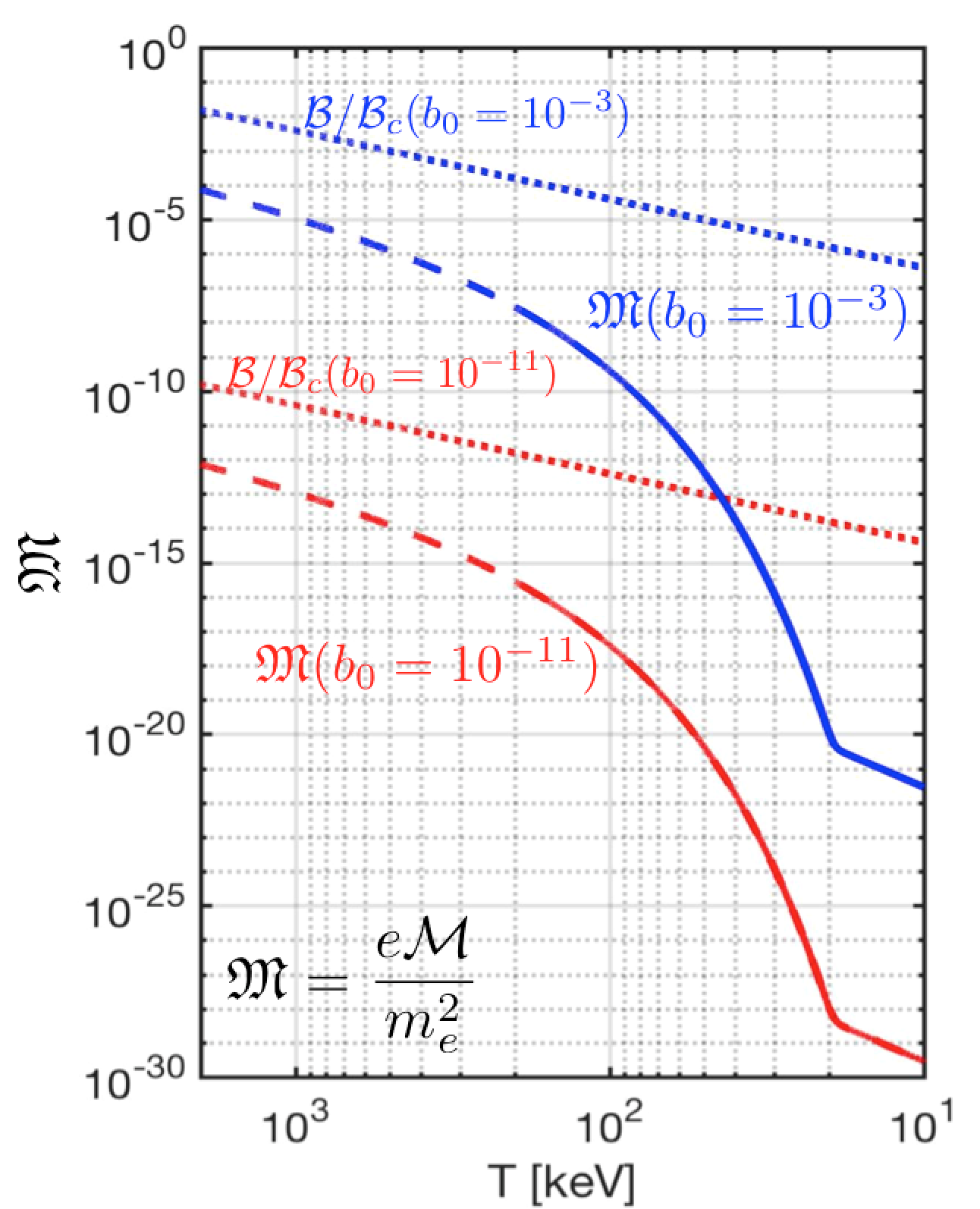}
 \caption{The magnetization ${\mathfrak M}$, with $g=2$, of the primordial $e^{+}e^{-}$-plasma is plotted as a function of temperature.}
 \label{fig:magnet} 
\end{figure}

\subsection{Magnetic response of electron-positron plasma}
\label{sec:paramagnetism}
\noindent In~\rf{fig:magnet}, we plot the magnetization as given by~\req{g2magplus} and~\req{g2magminus} with the magnetic moment potential set to unity $\xi=1$. The lower (solid red) and upper (solid blue) bounds for cosmic magnetic scale $b_{0}$ are included. The external magnetic field strength ${\cal B}/{\cal B}_{C}$ is also plotted for lower (dotted red) and upper (dotted blue) bounds. Since the derivative of the partition function governing magnetization may manifest differences between Fermi-Dirac and the here used Boltzmann limit more acutely, out of abundance of caution, we indicate extrapolation outside the domain of validity of the Boltzmann limit with dashes.

We see in~\rf{fig:magnet} that the $e^{+}e^{-}$-plasma is overall paramagnetic and yields a positive overall magnetization which is contrary to the traditional assumption that matter-antimatter plasma lack significant magnetic responses of their own in the bulk. With that said, the magnetization never exceeds the external field under the parameters considered which shows a lack of ferromagnetic behavior. 

The large abundance of pairs causes the smallness of the chemical potential seen in~\rf{fig:chemicalpotential} at high temperatures. As the universe expands and temperature decreases, there is a rapid decrease of the density $n_{e^{\pm}}$ of $e^{+}e^{-}$-pairs. This is the primary the cause of the rapid paramagnetic decrease seen in~\rf{fig:magnet} above $T_\mathrm{split}=20.3\keV$. At lower temperatures $T<20.3\keV$ there remains a small electron excess (see~\rf{fig:densityratio}) needed to neutralize proton charge. These excess electrons then govern the residual magnetization and dilutes with cosmic expansion.

An interesting feature of~\rf{fig:magnet} is that the magnetization in the full temperature range increases as a function of temperature. This is contrary to Curie's law~\cite{greiner2012thermodynamics} which stipulates that paramagnetic susceptibility of a laboratory material is inversely proportional to temperature. However, Curie's law applies to systems with fixed number of particles which is not true in our situation; see~\rsec{sec:perlepton}.

A further consideration is possible hysteresis as the $e^{+}e^{-}$ density drops with temperature. It is not immediately obvious the gas's magnetization should simply \lq degauss\rq\ so rapidly without further consequence. If the very large paramagnetic susceptibility present for $T\simeq m_{e}$ is the origin of an overall magnetization of the plasma, the conservation of magnetic flux through the comoving surface ensures that the initial residual magnetization is preserved at a lower temperature by Faraday induced kinetic flow processes however our model presented here cannot account for such effects. Some consequences of enforced magnetization are considered in~\rsec{sec:ferro}.

Early universe conditions may also apply to some extreme stellar objects with rapid change in $n_{e^{\pm}}$ with temperatures above $T_\mathrm{split}=20.3\keV$. Production and annihilation of $e^{+}e^{-}$-plasmas is also predicted around compact stellar objects~\cite{Ruffini:2009hg,Ruffini:2012it} potentially as a source of gamma-ray bursts (GRB).

\begin{figure}[ht]
 \centering
 \includegraphics[width=0.45\textwidth]{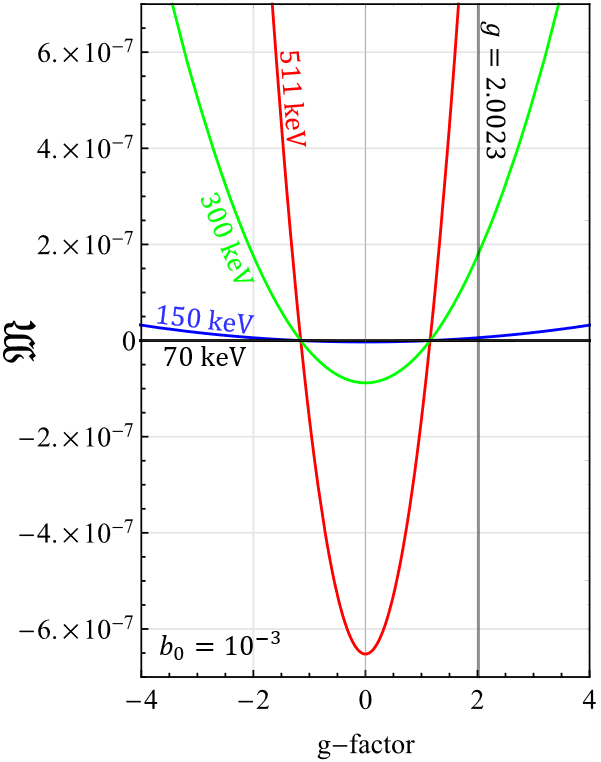}
 \caption{The magnetization $\mathfrak M$ as a function of $g$-factor plotted for several temperatures with magnetic scale $b_{0}=10^{-3}$ and polarization fugacity $\xi=1$.}
 \label{fig:gfac} 
\end{figure}

\subsection{g-factor balance between para and diamagnetism}
\label{sec:gfac}

\noindent As discussed at the end of~\rsec{sec:magnetization}, the AMM of $e^{+}e^{-}$ is not relevant in the present model. However out of academic interest, it is valuable to consider how magnetization is effected by changing the $g$-factor significantly.

The influence of AMM would be more relevant for the magnetization of baryon gasses since the $g$-factor for protons $(g\approx5.6)$ and neutrons $(g\approx3.8)$ are substantially different from $g=2$. The influence of AMM on the magnetization of thermal systems with large baryon content (neutron stars, magnetars, hypothetical bose stars, etc.) is therefore also of interest~\cite{Ferrer:2019xlr,Ferrer:2023pgq}.

\req{g2magplus} and~\req{g2magminus} with arbitrary $g$ reintroduced is given by
\begin{align}
\begin{split}
\label{arbg:1}
{\mathfrak M}&=\frac{e^{2}}{\pi^{2}}\frac{T^{2}}{m_{e}^{2}}\sum_{s'}^{\pm1}\xi_{s'}\cosh{\frac{\mu}{T}}\\
&\left[C^{1}_{s'}(x_{s'})K_{1}(x_{s'})+C^{0}_{s'}K_{0}(x_{s'})\right]\,,
\end{split}\\
\label{arbg:2}
C^{1}_{s'}(x_{\pm}) &= \left[\frac{1}{2}-\left(\frac{1}{2}-\frac{g}{4}s'\right)\left(1+\frac{b^2_0}{12x^{2}_{s'}}\right)\right]x_{s'}\,,\\
C^{0}_{s'} &= \left[\frac{1}{6}-\left(\frac{1}{4}-\frac{g}{8}s'\right)\right]b_0\,.
\end{align}
where $x_{s'}$ was previously defined in~\req{xfunc}.

In~\rf{fig:gfac}, we plot the magnetization as a function of $g$-factor between $4>g>-4$ for temperatures $T=\{511,\ 300,\ 150,\ 70\}\keV$. We find that the magnetization is sensitive to the value of AMM revealing a transition point between paramagnetic $({\mathfrak M}>0)$ and diamagnetic gasses $({\mathfrak M}<0)$. 

Curiously, the transition point was numerically determined to be around $g\simeq1.1547$ in the limit $b_{0}\rightarrow0$. The exact position of this transition point however was found to be both temperature and $b_{0}$ sensitive, though it moved little in the ranges considered.

It is not surprising for there to be a transition between diamagnetism and paramagnetism given that the partition function (see~\req{spin} and~\req{spinorbit}) contained elements of both. With that said, the transition point presented at $g\approx1.15$ should not be taken as exact because of the approximations used to obtain the above results. 

It is likely that the exact transition point has been altered by our taking of the Boltzmann approximation and Euler-Maclaurin integration steps. It is known that the Klein-Gordon-Pauli solutions to the Landau problem in~\req{kgp} have periodic behavior~\cite{Steinmetz:2018ryf,Evans:2022fsu,Rafelski:2022bsv} for $|g|=k/2$ (where $k\in1,2,3\ldots$).

These integer and half-integer points represent when the two Landau towers of orbital levels match up exactly. Therefore, we propose a more natural transition between the spinless diamagnetic gas of $g=0$ and a paramagnetic gas is $g=1$. A more careful analysis is required to confirm this, but that our numerical value is close to unity is suggestive.

\subsection{Laboratory versus the relativistic electron-positron-universe}
\label{sec:perlepton}
\noindent Despite the relatively large magnetization seen in~\rf{fig:magnet}, the average contribution per lepton is only a small fraction of its overall magnetic moment indicating the magnetization is only loosely organized. Specifically, the magnetization regime we are in is described by
\begin{align}
 \label{fractionalmagnetization}
 {\cal M}\ll\mu_{B}\frac{N_{e^{+}}+N_{e^{-}}}{V}\,,\qquad\mu_{B}\equiv\frac{e}{2m_{e}}\,,
\end{align}
where $\mu_{B}$ is the Bohr magneton and $N=nV$ is the total particle number in the proper volume V. To better demonstrate that the plasma is only weakly magnetized, we define the average magnetic moment per lepton given by along the field ($z$-direction) axis as
\begin{align}
 \label{momentperlepton}
 \vert\vec{m}\vert_{z}\equiv\frac{{\cal M}}{n_{e^{-}}+n_{e^{+}}}\,,\qquad\vert\vec{m}\vert_{x}=\vert\vec{m}\vert_{y}=0\,.
\end{align}
Statistically, we expect the transverse expectation values to be zero. We emphasize here that despite $|\vec{m}|_{z}$ being nonzero, this doesn't indicate a nonzero spin angular momentum as our plasma is nearly matter-antimatter symmetric. The quantity defined in~\req{momentperlepton} gives us an insight into the microscopic response of the plasma.

\begin{figure}[ht]
 \centering
 \includegraphics[width=0.45\textwidth]{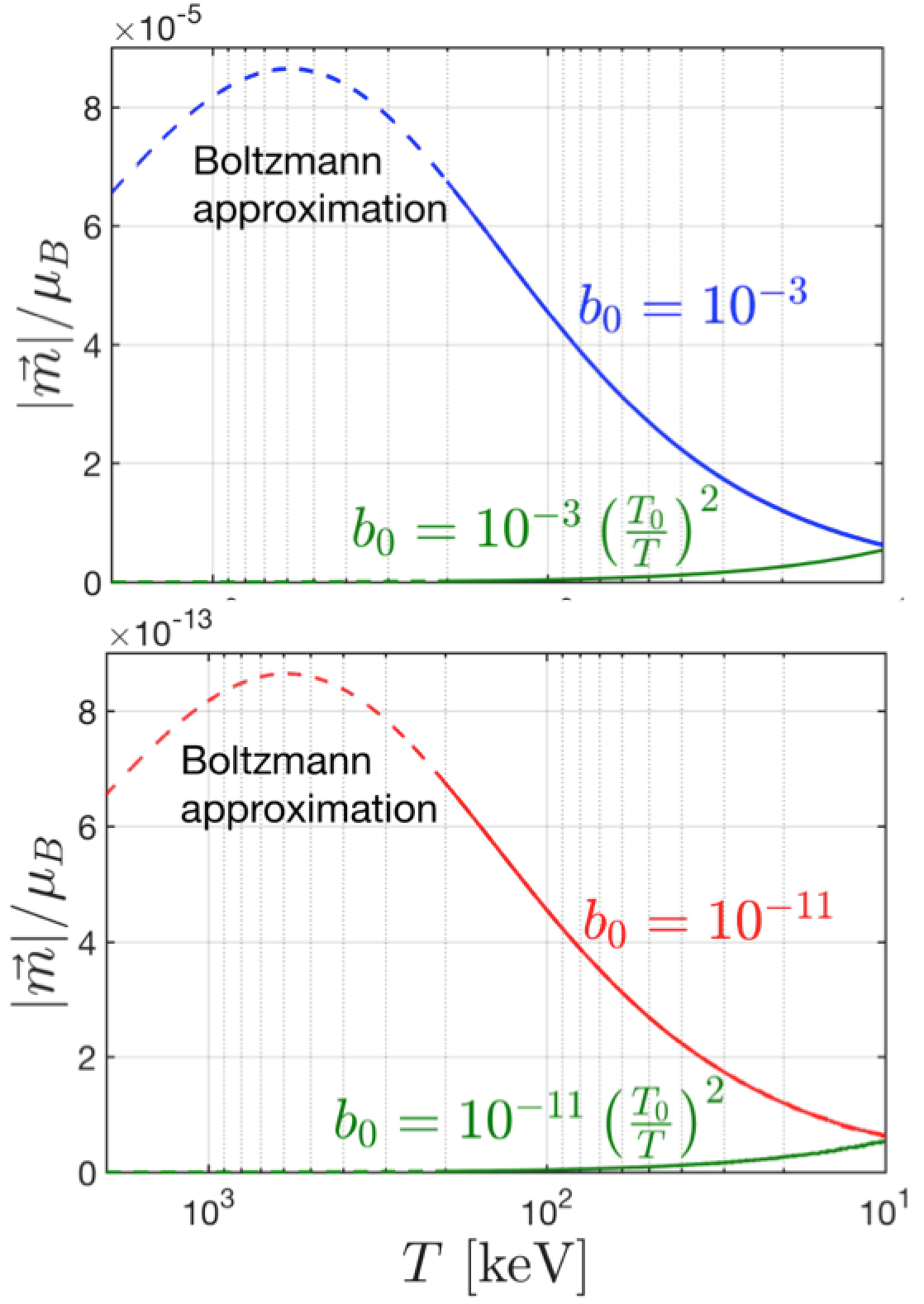}
 \caption{The magnetic moment per lepton $\vert\vec{m}\vert_{z}$ along the field axis as a function of temperature.}
 \label{fig:momentperlepton}
\end{figure}

The average magnetic moment $\vert\vec{m}\vert_{z}$ defined in~\req{momentperlepton} is plotted in~\rf{fig:momentperlepton} which displays how essential the external field is on the \lq per lepton\rq\ magnetization. Both the $b_{0}=10^{-11}$ (lower plot, red curve) and $b_{0}=10^{-3}$ (upper plot, blue curve) cosmic magnetic scale bounds are plotted in the Boltzmann approximation. The dashed lines indicate where this approximation is only qualitatively correct. For illustration, a constant magnetic field case (solid green line) with a comoving reference value chosen at temperature $T_{0}=10\keV$ is also plotted.

If the field strength is held constant, then the average magnetic moment per lepton is suppressed at higher temperatures as expected for magnetization satisfying Curie's law. The difference in~\rf{fig:momentperlepton} between the non-constant (red and blue solid curves) case and the constant field (solid green curve) case demonstrates the importance of the conservation of primordial magnetic flux in the plasma, required by~\req{bscale}.

While not shown, if~\rf{fig:momentperlepton} was extended to lower temperatures, the magnetization per lepton of the constant field case would be greater than the non-constant case which agrees with our intuition that magnetization is easier to achieve at lower temperatures. This feature again highlights the importance of flux conservation in the system and the uniqueness of the primordial cosmic environment.

\section{Magnetic moment polarization and ferromagnetism}
\label{sec:ferro}
\subsection{Magnetic moment chemical potential}
\label{sec:spinpot}
\noindent Up to this point, we have neglected the impact that a nonzero magnetic moment potential $\eta\neq0$ (and thus $\xi\neq1$) would have on the primordial $e^{+}e^{-}$-plasma magnetization. In the limit that $(m_{e}/T)^2\gg b_0$ the magnetization given in~\req{arbg:1} and~\req{arbg:2} is entirely controlled by the magnetic moment fugacity $\xi$ asymmetry generated by the magnetic moment potential $\eta$ yielding up to first order ${\cal O}(b_{0})$ in magnetic scale
\begin{multline}
 \label{ferro}
 \lim_{m_{e}^{2}/T^{2}\gg b_0}{\mathfrak M}=\frac{g}{2}\frac{e^{2}}{\pi^{2}}\frac{T^{2}}{m_{e}^{2}}\sinh{\frac{\eta}{T}}\cosh{\frac{\mu}{T}}\left[\frac{m_{e}}{T}K_{1}\left(\frac{m_{e}}{T}\right)\right]\\
 +b_{0}\left(g^{2}-\frac{4}{3}\right)\frac{e^{2}}{8\pi^{2}}\frac{T^{2}}{m_{e}^{2}}\cosh{\frac{\eta}{T}}\cosh{\frac{\mu}{T}}K_{0}\left(\frac{m_{e}}{T}\right)
 +{\cal O}\left(b_{0}^{2}\right)
\end{multline}

Given~\req{ferro}, we can understand the magnetic moment potential as a kind of \lq ferromagnetic\rq\ influence on the primordial gas which allows for magnetization even in the absence of external magnetic fields. This interpretation is reinforced by the fact the leading coefficient is $g/2$.

We suggest that a variety of physics could produce a small nonzero $\eta$ within a domain of the gas. Such asymmetries could also originate statistically as while the expectation value of free gas polarization is zero, the variance is likely not.

As $\sinh{\eta/T}$ is an odd function, the sign of $\eta$ also controls the alignment of the magnetization. In the high temperature limit~\req{ferro} with strictly $b_{0}=0$ assumes a form of to lowest order for brevity
\begin{align}
 \label{hiTferro}
 \lim_{m_{e}/T\rightarrow0}{\mathfrak M}\vert_{b_{0}=0}=\frac{g}{2}\frac{e^{2}}{\pi^{2}}\frac{T^{2}}{m_{e}^{2}}\frac{\eta}{T}\,,
\end{align}

While the limit in~\req{hiTferro} was calculated in only the Boltzmann limit, it is noteworthy that the high temperature (and $m\rightarrow0$) limit of Fermi-Dirac distributions only differs from the Boltzmann result by a proportionality factor. 

The natural scale of the $e^{+}e^{-}$ magnetization with only a small magnetic moment fugacity ($\eta<1\eV$) fits easily within the bounds of the predicted magnetization during this era if the IGMF measured today was of primordial origin. The reason for this is that the magnetization seen in~\req{g2magplus},~\req{g2magminus} and~\req{ferro} are scaled by $\alpha{\cal B}_{C}$ where $\alpha$ is the fine structure constant.

\subsection{Self-magnetization}
\label{sec:self}

\noindent One exploratory model we propose is to fix the magnetic moment polarization asymmetry, described in~\req{spotential}, to generate a homogeneous magnetic field which dissipates as the universe cools down. In this model, there is no pre-existing external primordial magnetic field generated by some unrelated physics, but rather the $e^{+}e^{-}$-plasma itself is responsible for the creation of $({\cal B}_{\rm PMF}\ne 0)$ field by virtue of magnetic moment polarization. 
\begin{figure}[ht]
 \centering
 \includegraphics[width=0.45\textwidth]{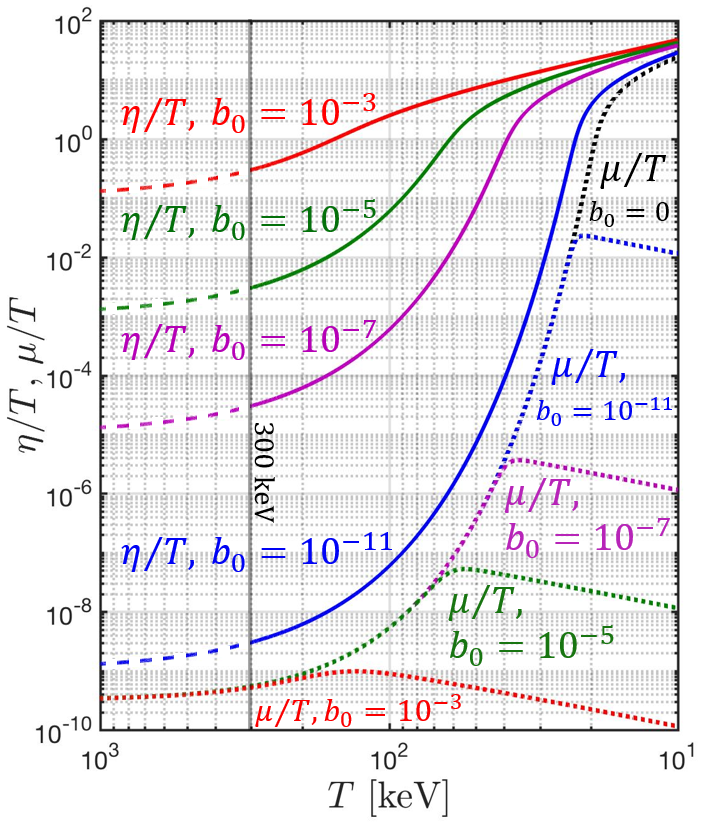}
 \caption{The magnetic moment potential $\eta$ and chemical potential $\mu$ are plotted under the assumption of self-magnetization through a nonzero magnetic moment polarization in bulk of the plasma.}
 \label{fig:self} 
\end{figure}

This would obey the following assumption of
\begin{align}
 \label{selfmag}
 {\mathfrak M}(b_{0})=\frac{{\cal M}(b_0)}{{\cal B}_{C}}\longleftrightarrow\frac{\cal B}{{\cal B}_{C}}=b_{0}\frac{T^{2}}{m_{e}^{2}}\,,
\end{align}
which sets the total magnetization as a function of itself. The magnetic moment polarization described by $\eta\rightarrow\eta(b_{0},T)$ then becomes a fixed function of the temperature and magnetic scale. The underlying assumption would be the preservation of the homogeneous field would be maintained by scattering within the gas (as it is still in thermal equilibrium) modulating the polarization to conserve total magnetic flux.

The result of the self-magnetization assumption in~\req{selfmag} for the potentials is plotted in~\rf{fig:self}. The solid lines indicate the curves for $\eta/T$ for differing values of $b_{0}=\{10^{-11},\ 10^{-7},\ 10^{-5},\ 10^{-3}\}$ which become dashed above $T=300\keV$ to indicate that the Boltzmann approximation is no longer appropriate though the general trend should remain unchanged. 

The dotted lines are the curves for the chemical potential $\mu/T$. At high temperatures we see that a relatively small $\eta/T$ is needed to produce magnetization owing to the large densities present.~\rf{fig:self} also shows that the chemical potential does not deviate from the free particle case until the magnetic moment polarization becomes sufficiently high which indicates that this form of self-magnetization would require the annihilation of positrons to be incomplete even at lower temperatures.

This is seen explicitly in~\rf{fig:polarswap} where we plot the numerical density of particles as a function of temperature for spin aligned $(+\eta)$ and spin anti-aligned $(-\eta)$ species for both positrons $(-\mu)$ and electrons $(+\mu)$. Various self-magnetization strengths are also plotted to match those seen in~\rf{fig:self}. The nature of $T_{\rm split}$ changes under this model since antimatter and polarization states can be extinguished separately. Positrons persist where there is insufficient electron density to maintain the magnetic flux. Polarization asymmetry therefore appears physical only in the domain where there is a large number of matter-antimatter pairs.

\begin{figure}
 \centering
\includegraphics[width=0.45\textwidth]{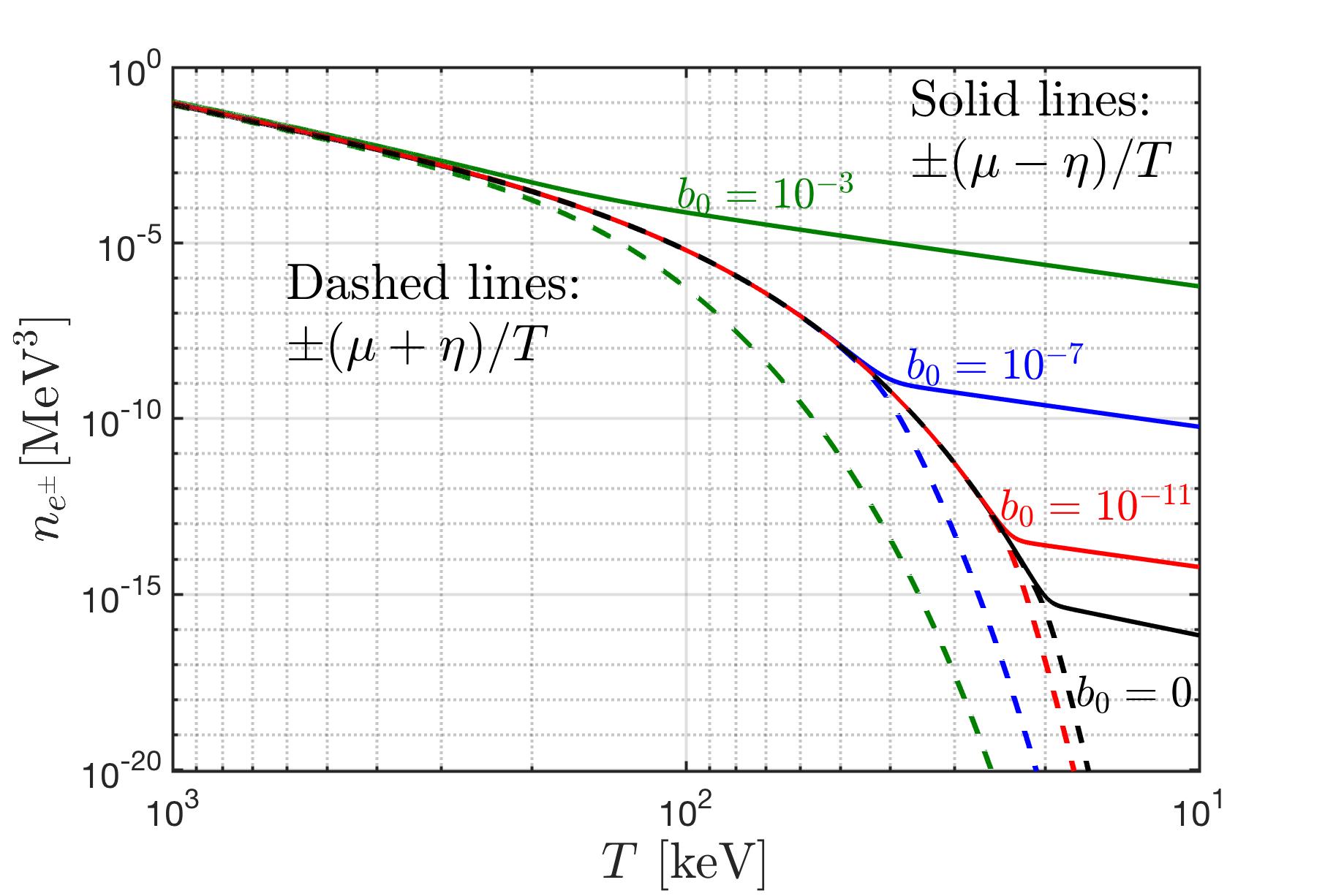}
 \caption{The number density $n_{e^{\pm}}$ of polarized electrons and positrons under the self-magnetization model for differing values of $b_{0}$.}
 \label{fig:polarswap} 
\end{figure}

The low $T$-behavior of~\rf{fig:polarswap} will need further corroboration after Amp{\'e}rian currents as a source of magnetic field are incorporated: The Gilbertian sources, here magnetic dipole moment paramagnetism and Landau diamagnetism, may not be dominant magnetic sources when $e^{+}e^{-}$-pairs are of comparable number to the residual electron and proton abundance.

\section{Summary and discussion}
\label{sec:conclusions}
\noindent This work is an effort to interpret the intergalactic magnetic fields of today as originating in primordial fields generated in the first hour of the universe's existence. In~\rsec{sec:abundance} have demonstrated that the $e^{+}e^{-}$-pair plasma is an appropriate non-electrical current candidate source for the primordial field: It is (a) very dense, (b) made of particles with highest magnetic moment in nature, (c) and displays a strong paramagnetic response. Therefore expanding on our work in~\cite{Rafelski:2023emw}, we explored its paramagnetic magnetization in the early universe temperature range between $2000\keV>T>20\keV$. 

In~\rsec{sec:thermal} we define the comoving scale, $b_{0}$, of the magnetic field expressed in dimensionless units estimated between $10^{-3}>b_{0}>10^{-11}$. We believe considering conservation of magnetic flux that $b_{0}$ once created is most likely a conserved property of the expanding universe. Antiparticles $(e^{+})$ have the opposite sign of charge, and thus magnetic moment, compared to particles $(e^{-})$. Therefore in an $e^{+}e^{-}$-pair plasma, net magnetization can be associated with opposite spin orientations for particles and antiparticles without the accompaniment of a net angular momentum in the volume considered. This is of course very different from the the matter dominated universe arising below $T\simeq 20\keV$ which includes the current epoch.  

The $e^{+}e^{-}$-pair plasma environment is well beyond the reach of all present day laboratory and known astrophysical environments. As seen in~\rf{fig:densityratio} the lower temperature limit, where the last $e^{+}e^{-}$-pair disappeared, is 15 times the Sun's core temperature~\cite{Bahcall:2000nu} $T_{\odot}=1.37\keV$. Laboratory conditions to explore our results depend on presence of $e^{+}e^{-}$-pair abundance which in turn depends on sufficiently stable thermal photon content.

Both Landau diamagnetism and magnetic dipole moment paramagnetism are relevant in the analysis of dense $e^{+}e^{-}$-plasma Gilbertian (non-current) magnetization. The high temperature relevance of paramagnetism relies on the high abundance of pairs. In the theoretical treatment of~\rsec{sec:thermal}, this is accounted for by introducing effective polarization mass $\tilde{m}$ in~\req{effmass:2}. This allows for the separation of the spin portion of the relativistic partition function from the spin-orbital portion (Landau diamagnetism) in~\rsec{sec:paradia}. In~\rsec{sec:chem} we determined the effect of magnetism on the chemical potential; 
see~\rf{fig:chemicalpotential}.

This novel approach to high temperature magnetization allows using~\rsec{sec:magnetization} to show that the $e^{+}e^{-}$-plasma paramagnetic response (see~\req{g2magplus} and~\req{g2magminus}) is dominated by the varying abundance of electron-positron pairs, decreasing with decreasing $T$ for $T<m_{e}c^2$. This is unlike conventional laboratory cases where the number of magnetic particles is constant. 

In our domain of interest, we determine in~\rsec{sec:gfac} that cosmic magnetization is not sensitive to the anomalous magnetic moment of the electron. Considering magnetization as a function of $g$-factor we find a transition seen in~\rf{fig:gfac} between paramagnetic and diamagnetic gasses. The per-lepton magnetization is shown in~\rsec{sec:perlepton} and~\rf{fig:momentperlepton} indicating the plasma is only weakly organized in its response.

In~\rsec{sec:self} we explored spin asymmetry (defined in~\rsec{sec:abundance} and~\req{spotential}) via the magnetic moment  chemical potential. We showed in~\rsec{sec:ferro} how self-magnetization can be induced by magnetic moment polarization via a novel magnetic moment fugacity. We obtained in~\rsec{sec:self} the required primordial degree of magnetic moment polarization necessary to understand today's IGMF. Our study demonstrates that the early universe required at high temperatures only a minute asymmetry in magnetic moment polarization to produce required spontaneous magnetization.

Our results lead to extensions of present day paradigms but offer many opportunities for improvement. The high temperature domain would require a full inventory of particles including neutrinos and muons before their disappearance or decoupling. The full Fermi-Dirac and Bose-Einstein statistics instead of the Boltzmann approximation would then be employed. 

Beyond $e^{+}e^{-}$-plasma, the quark-gluon plasma at $T>150\,000\keV$ is also of great interest. The up-quark has the largest natural charge-to-mass $e/m$ ratio among elementary particles besides the electron. A connection from the quark-gluon plasma to the $e^{+}e^{-}$-plasma then requires understanding of the impact of the hadronization process on magnetization, and vice-verse, a consideration of hadronization as a magnetization mechanism. We note that the contribution of $\mu^{+}\mu^{-}$-plasma to magnetization is reduced by a factor $\simeq 200$ compared to $e^{+}e^{-}$-plasma due to the $\propto e/m_{\mu}$ behavior of magnetic moments. We also note that the complex neutrino decoupling process near to $T=2\,000\keV$ should be explored as a source of magnetization mechanism.
 
Near to $T=80\keV$ just prior to BBN we have $4.47\times10^{8}$ $e^{+}e^{-}$-pairs per baryon and a primordial magnetic field in range of $10^{9}-10^{1}$ Gauss. BBN thus occurs in an environment as different as can be imagined from the empty space network of nuclear reactions explored. Our work creates the question in what way the presence of a primordial magnetic field could have impacted BBN and vice-verse, if BBN could provide the mechanism for spontaneous magnetization. The $e^{+}e^{-}$-pair impact is already being considered~\cite{Grayson:2023flr}.

Below $T_\mathrm{split}=20.3\keV$ the universe's particle inventory is dominated by electrons, protons, and $\alpha$-particles. In order to conserve the magnetic flux originating in the polarized homogeneous $e^{+}e^{-}$-pair plasma a very different model will need to be developed allowing for fragmentation of the homogeneous plasma universe into polarization domains and evolving Amp{\`e}rian kinetic current curl responses. The effort to connect the bulk magnetization due to discrete dipoles by Amp{\`e}rian magnetization generated through currents and inhomogeneous flows will require study of transport equations allowing for magnetic moment polarization. 

Recent measurements by the James Webb Space Telescope (JWST)~\cite{Yan:2022sxd,adams2023discovery,arrabal2023spectroscopic} indicate that maturing galaxies already present at a large redshift value of $z\gtrsim10$ within the first 500 million years of the universe. This requires gravitational collapse to begin earlier in a hotter environment. Additionally the observation of supermassive (with millions of solar masses) black holes already present~\cite{CEERSTeam:2023qgy} in this same high redshift era indicate the need for exceptionally small-scale high mass density regions in the early universe. There is a natural mechanism present in our work needed to create the above condition: As the universe evolved, the rapid $10^{8}$ drop in $e^{+}e^{-}$ abundance within the temperature range $100\keV>T>20.3\keV$ shown in~\rf{fig:densityratio} could be inducing dynamical currents preserving (comoving) magnetic flux in the emerging $p^{+}\alpha^{++}e^{-}$-plasma and in turn generate vortex seeds for small scale baryonic matter localization which could support anisotropies in the cosmic microwave background (CMB)~\cite{Jedamzik:2013gua,Abdalla:2022yfr}.

To conclude: This work shows that the paramagnetic and diamagnetic $e^{+}e^{-}$-plasma properties may play a pivotal role in understanding the primordial universe. In particular we have shown that the possible self-magnetization of the cosmic $e^{+}e^{-}$-plasma  provides a novel and credible proposal for interpretation and exploration of magnetic fields in the universe.

\acknowledgments
\label{sec:ack}
\noindent Johann Rafelski would like to acknowledge the fruitful discussions with Massimo Giovannini at CERN which partly inspired this work.

\bibliographystyle{unsrtnat}
\bibliography{refs-plasma-partition}

\begin{thebibliography}{48}
\providecommand{\natexlab}[1]{#1}
\providecommand{\url}[1]{\texttt{#1}}
\expandafter\ifx\csname urlstyle\endcsname\relax
  \providecommand{\doi}[1]{doi: #1}\else
  \providecommand{\doi}{doi: \begingroup \urlstyle{rm}\Url}\fi

\bibitem[Neronov and Vovk(2010)]{Neronov:2010gir}
A.~Neronov and I.~Vovk.
\newblock Evidence for strong extragalactic magnetic fields from fermi
  observations of tev blazars.
\newblock \emph{Science}, 328\penalty0 (5974):\penalty0 73--75, 2010.
\newblock \doi{10.1126/science.1184192}.

\bibitem[Taylor et~al.(2011)Taylor, Vovk, and Neronov]{Taylor:2011bn}
A.~M. Taylor, I.~Vovk, and A.~Neronov.
\newblock Extragalactic magnetic fields constraints from simultaneous gev--tev
  observations of blazars.
\newblock \emph{Astronomy \& Astrophysics}, 529:\penalty0 A144, 2011.
\newblock \doi{10.1051/0004-6361/201116441}.

\bibitem[Pshirkov et~al.(2016)Pshirkov, Tinyakov, and Urban]{Pshirkov:2015tua}
M.~S. Pshirkov, P.~G. Tinyakov, and F.~R. Urban.
\newblock {New limits on extragalactic magnetic fields from rotation measures}.
\newblock \emph{Phys. Rev. Lett.}, 116\penalty0 (19):\penalty0 191302, 2016.
\newblock \doi{10.1103/PhysRevLett.116.191302}.

\bibitem[Jedamzik and Saveliev(2019)]{Jedamzik:2018itu}
K.~Jedamzik and A.~Saveliev.
\newblock Stringent limit on primordial magnetic fields from the cosmic
  microwave background radiation.
\newblock \emph{Physical review letters}, 123\penalty0 (2):\penalty0 021301,
  2019.
\newblock \doi{10.1103/PhysRevLett.123.021301}.

\bibitem[Vernstrom et~al.(2021)Vernstrom, Heald, Vazza, Galvin, West,
  Locatelli, Fornengo, and Pinetti]{Vernstrom:2021hru}
T.~Vernstrom, G.~Heald, F.~Vazza, T.~J. Galvin, J.~L. West, N.~Locatelli,
  N.~Fornengo, and E.~Pinetti.
\newblock {Discovery of magnetic fields along stacked cosmic filaments as
  revealed by radio and X-ray emission}.
\newblock \emph{Monthly Notices of the Royal Astronomical Society},
  505\penalty0 (3):\penalty0 4178--4196, 05 2021.
\newblock \doi{10.1093/mnras/stab1301}.

\bibitem[Giovannini(2018)]{Giovannini:2017rbc}
M.~Giovannini.
\newblock {Probing large-scale magnetism with the Cosmic Microwave Background}.
\newblock \emph{Class. Quant. Grav.}, 35\penalty0 (8):\penalty0 084003, 2018.
\newblock \doi{10.1088/1361-6382/aab17d}.

\bibitem[Giovannini(2004)]{Giovannini:2003yn}
M.~Giovannini.
\newblock {The Magnetized universe}.
\newblock \emph{Int. J. Mod. Phys. D}, 13:\penalty0 391--502, 2004.
\newblock \doi{10.1142/S0218271804004530}.

\bibitem[Kronberg(1994)]{Kronberg:1993vk}
P.~P. Kronberg.
\newblock Extragalactic magnetic fields.
\newblock \emph{Reports on Progress in Physics}, 57\penalty0 (4):\penalty0 325,
  1994.
\newblock \doi{10.1088/0034-4885/57/4/001}.

\bibitem[Rafelski et~al.(2023{\natexlab{a}})Rafelski, Birrell, Steinmetz, and
  Yang]{Rafelski:2023emw}
J.~Rafelski, J.~Birrell, A.~Steinmetz, and C.~T. Yang.
\newblock {A Short Survey of Matter-Antimatter Evolution in the Primordial
  Universe}.
\newblock \emph{Universe}, 9\penalty0 (7):\penalty0 309, 2023{\natexlab{a}}.
\newblock \doi{10.3390/universe9070309}.

\bibitem[Grayson et~al.(2023)Grayson, Yang, Formanek, and
  Rafelski]{Grayson:2023flr}
C.~Grayson, C.~T. Yang, M.~Formanek, and J.~Rafelski.
\newblock Electron-positron plasma in \uppercase{BBN}: damped-dynamic
  screening.
\newblock \emph{arXiv preprint}, 2023.
\newblock \doi{10.48550/arXiv.2307.11264}.
\newblock [in press in Annals of Physics].

\bibitem[Rafelski et~al.(2018)Rafelski, Formanek, and
  Steinmetz]{Rafelski:2017hce}
J.~Rafelski, M.~Formanek, and A.~Steinmetz.
\newblock {Relativistic Dynamics of Point Magnetic Moment}.
\newblock \emph{Eur. Phys. J. C}, 78\penalty0 (1):\penalty0 6, 2018.
\newblock \doi{10.1140/epjc/s10052-017-5493-2}.

\bibitem[Giovannini(2023)]{Giovannini:2022rrl}
M.~Giovannini.
\newblock The scaling of primordial gauge fields.
\newblock \emph{Physics Letters B}, 842:\penalty0 137967, 2023.
\newblock ISSN 0370-2693.
\newblock \doi{10.1016/j.physletb.2023.137967}.

\bibitem[Batista and Saveliev(2021)]{AlvesBatista:2021sln}
R.~A. Batista and A.~Saveliev.
\newblock The gamma-ray window to intergalactic magnetism.
\newblock \emph{Universe}, 7\penalty0 (7), 2021.
\newblock ISSN 2218-1997.
\newblock \doi{10.3390/universe7070223}.

\bibitem[Gaensler et~al.(2004)Gaensler, Beck, and Feretti]{Gaensler:2004gk}
B.~M. Gaensler, R.~Beck, and L.~Feretti.
\newblock The origin and evolution of cosmic magnetism.
\newblock \emph{New Astronomy Reviews}, 48\penalty0 (11-12):\penalty0
  1003--1012, 2004.
\newblock \doi{10.1016/j.newar.2004.09.003}.

\bibitem[Durrer and Neronov(2013)]{Durrer:2013pga}
R.~Durrer and A.~Neronov.
\newblock Cosmological magnetic fields: their generation, evolution and
  observation.
\newblock \emph{The Astronomy and Astrophysics Review}, 21:\penalty0 1--109,
  2013.
\newblock \doi{10.1007/s00159-013-0062-7}.

\bibitem[{Pomakov} et~al.(2022){Pomakov}, {O'Sullivan}, {Br{\"u}ggen}, {Vazza},
  {Carretti}, {Heald}, {Horellou}, {Shimwell}, {Shulevski}, and
  {Vernstrom}]{Pomakov:2022cem}
V.~P. {Pomakov}, S.~P. {O'Sullivan}, M.~{Br{\"u}ggen}, F.~{Vazza},
  E.~{Carretti}, G.~H. {Heald}, C.~{Horellou}, T.~{Shimwell}, A.~{Shulevski},
  and T.~{Vernstrom}.
\newblock {The redshift evolution of extragalactic magnetic fields}.
\newblock \emph{Monthly Notices of the Royal Astronomical Society},
  515\penalty0 (1):\penalty0 256--270, 2022.
\newblock \doi{10.1093/mnras/stac1805}.

\bibitem[Jedamzik and Pogosian(2020)]{Jedamzik:2020krr}
K.~Jedamzik and L.~Pogosian.
\newblock Relieving the hubble tension with primordial magnetic fields.
\newblock \emph{Physical Review Letters}, 125\penalty0 (18):\penalty0 181302,
  2020.
\newblock \doi{10.1103/PhysRevLett.125.181302}.

\bibitem[Birrell et~al.(2014)Birrell, Yang, and Rafelski]{Birrell:2014uka}
J.~Birrell, C.~T. Yang, and J.~Rafelski.
\newblock {Relic Neutrino Freeze-out: Dependence on Natural Constants}.
\newblock \emph{Nucl. Phys. B}, 890:\penalty0 481--517, 2014.
\newblock \doi{10.1016/j.nuclphysb.2014.11.020}.

\bibitem[Bahcall et~al.(2001)Bahcall, Pinsonneault, and Basu]{Bahcall:2000nu}
J.~N. Bahcall, M.~H. Pinsonneault, and S.~Basu.
\newblock Solar models: Current epoch and time dependences, neutrinos, and
  helioseismological properties.
\newblock \emph{The Astrophysical Journal}, 555\penalty0 (2):\penalty0 990, jul
  2001.
\newblock \doi{10.1086/321493}.

\bibitem[Fromerth et~al.(2012)Fromerth, Kuznetsova, Labun, Letessier, and
  Rafelski]{Fromerth:2012fe}
M.~J. Fromerth, I.~Kuznetsova, L.~Labun, J.~Letessier, and J.~Rafelski.
\newblock {From Quark-Gluon Universe to Neutrino Decoupling: 200 \ensuremath{<}
  T \ensuremath{<} 2MeV}.
\newblock \emph{Acta Phys. Polon. B}, 43\penalty0 (12):\penalty0 2261--2284,
  2012.
\newblock \doi{10.5506/APhysPolB.43.2261}.

\bibitem[Canetti et~al.(2012)Canetti, Drewes, and Shaposhnikov]{Canetti:2012zc}
L.~Canetti, M.~Drewes, and M.~Shaposhnikov.
\newblock {Matter and Antimatter in the Universe}.
\newblock \emph{New J. Phys.}, 14:\penalty0 095012, 2012.
\newblock \doi{10.1088/1367-2630/14/9/095012}.

\bibitem[Elze et~al.(1980)Elze, Greiner, and Rafelski]{Elze:1980er}
H.~T. Elze, W.~Greiner, and J.~Rafelski.
\newblock {The relativistic Fermi gas revisited}.
\newblock \emph{J. Phys. G}, 6:\penalty0 L149--L153, 1980.
\newblock \doi{10.1088/0305-4616/6/9/003}.

\bibitem[Workman et~al.(2022)]{ParticleDataGroup:2022pth}
R.~L. Workman et~al.
\newblock {Review of Particle Physics}.
\newblock \emph{PTEP}, 2022:\penalty0 083C01, 2022.
\newblock \doi{10.1093/ptep/ptac097}.

\bibitem[Letessier and Rafelski(2023)]{Letessier:2002ony}
J.~Letessier and J.~Rafelski.
\newblock \emph{Hadrons and Quark–Gluon Plasma}.
\newblock Cambridge Monographs on Particle Physics, Nuclear Physics and
  Cosmology. Cambridge University Press, 2023.
\newblock \doi{10.1017/9781009290753}.
\newblock \emph{Open access.} [Orig. pub. year: 2002].

\bibitem[Abdalla et~al.(2022)Abdalla, Abell{\'a}n, Aboubrahim, Agnello, Akarsu,
  Akrami, Alestas, Aloni, Amendola, Anchordoqui, et~al.]{Abdalla:2022yfr}
E.~Abdalla, G.~F. Abell{\'a}n, A.~Aboubrahim, A.~Agnello, {\"O}.~Akarsu,
  Y.~Akrami, G.~Alestas, D.~Aloni, L.~Amendola, L.~A. Anchordoqui, et~al.
\newblock Cosmology intertwined: A review of the particle physics,
  astrophysics, and cosmology associated with the cosmological tensions and
  anomalies.
\newblock \emph{Journal of High Energy Astrophysics}, 34:\penalty0 49--211,
  2022.
\newblock \doi{10.1016/j.jheap.2022.04.002}.

\bibitem[Weinberg(1972)]{weinberg1972gravitation}
S.~Weinberg.
\newblock \emph{Gravitation and cosmology: principles and applications of the
  general theory of relativity}.
\newblock John Wiley \& Sons, 1972.

\bibitem[Aghanim et~al.(2020)]{Planck:2018vyg}
N.~Aghanim et~al.
\newblock {Planck 2018 results. VI. Cosmological parameters}.
\newblock \emph{Astron. Astrophys.}, 641:\penalty0 A6, 2020.
\newblock \doi{10.1051/0004-6361/201833910}.
\newblock [Erratum: Astron.Astrophys. 652, C4 (2021)].

\bibitem[Melrose(2013)]{melrose2008quantum}
D.~Melrose.
\newblock \emph{Quantum plasmadynamics: Magnetized plasmas}.
\newblock Springer, 2013.
\newblock \doi{10.1007/978-1-4614-4045-1}.

\bibitem[Steinmetz et~al.(2019)Steinmetz, Formanek, and
  Rafelski]{Steinmetz:2018ryf}
A.~Steinmetz, M.~Formanek, and J.~Rafelski.
\newblock {Magnetic Dipole Moment in Relativistic Quantum Mechanics}.
\newblock \emph{Eur. Phys. J. A}, 55\penalty0 (3):\penalty0 40, 2019.
\newblock \doi{10.1140/epja/i2019-12715-5}.

\bibitem[Tiesinga et~al.(2021)Tiesinga, Mohr, Newell, and
  Taylor]{Tiesinga:2021myr}
Eite Tiesinga, Peter~J. Mohr, David~B. Newell, and Barry~N. Taylor.
\newblock {CODATA recommended values of the fundamental physical constants:
  2018}.
\newblock \emph{Rev. Mod. Phys.}, 93\penalty0 (2):\penalty0 025010, 2021.
\newblock \doi{10.1103/RevModPhys.93.025010}.

\bibitem[Abramowitz et~al.(1988)Abramowitz, Stegun, and
  Romer]{abramowitz1988handbook}
M.~Abramowitz, I.~A. Stegun, and R.~H. Romer.
\newblock \emph{Handbook of mathematical functions with formulas, graphs, and
  mathematical tables}.
\newblock American Association of Physics Teachers, 1988.

\bibitem[Greiner et~al.(2012)Greiner, Neise, and
  St{\"o}cker]{greiner2012thermodynamics}
W.~Greiner, L.~Neise, and H.~St{\"o}cker.
\newblock \emph{Thermodynamics and statistical mechanics}.
\newblock Springer Science \& Business Media, 2012.
\newblock \doi{10.1007/978-1-4612-0827-3}.
\newblock [Orig. pub. year: 1995].

\bibitem[Greiner and Reinhardt(2008)]{greiner2008quantum}
W.~Greiner and J.~Reinhardt.
\newblock \emph{Quantum electrodynamics}.
\newblock Springer Science \& Business Media, 2008.
\newblock \doi{10.1007/978-3-540-87561-1}.

\bibitem[Vazza et~al.(2017)Vazza, Br\"uggen, Gheller, Hackstein, Wittor, and
  Hinz]{Vazza:2017qge}
F.~Vazza, M.~Br\"uggen, C.~Gheller, S.~Hackstein, D.~Wittor, and P.~M. Hinz.
\newblock {Simulations of extragalactic magnetic fields and of their
  observables}.
\newblock \emph{Class. Quant. Grav.}, 34\penalty0 (23):\penalty0 234001, 2017.
\newblock \doi{10.1088/1361-6382/aa8e60}.

\bibitem[Vachaspati(2021)]{Vachaspati:2020blt}
T.~Vachaspati.
\newblock {Progress on cosmological magnetic fields}.
\newblock \emph{Rept. Prog. Phys.}, 84\penalty0 (7):\penalty0 074901, 2021.
\newblock \doi{10.1088/1361-6633/ac03a9}.

\bibitem[Gopal and Sethi(2005)]{Gopal:2004ut}
R.~Gopal and S.~Sethi.
\newblock {Generation of magnetic field in the pre-recombination era}.
\newblock \emph{Mon. Not. Roy. Astron. Soc.}, 363:\penalty0 521--528, 2005.
\newblock \doi{10.1111/j.1365-2966.2005.09442.x}.

\bibitem[Perrone et~al.(2021)Perrone, Gregori, Reville, Silva, and
  Bingham]{Perrone:2021srr}
L.~M. Perrone, G.~Gregori, B.~Reville, L.~O. Silva, and R.~Bingham.
\newblock {Neutrino-electron magnetohydrodynamics in an expanding universe}.
\newblock \emph{Phys. Rev. D}, 104\penalty0 (12):\penalty0 123013, 2021.
\newblock \doi{10.1103/PhysRevD.104.123013}.

\bibitem[Evans and Rafelski(2022)]{Evans:2022fsu}
S.~Evans and J.~Rafelski.
\newblock {Emergence of periodic in magnetic moment effective QED action}.
\newblock \emph{Phys. Lett. B}, 831:\penalty0 137190, 2022.
\newblock \doi{10.1016/j.physletb.2022.137190}.

\bibitem[Rafelski et~al.(2023{\natexlab{b}})Rafelski, Evans, and
  Labun]{Rafelski:2022bsv}
J.~Rafelski, S.~Evans, and L.~Labun.
\newblock {Study of QED singular properties for variable gyromagnetic ratio
  $g\simeq 2$}.
\newblock \emph{Phys. Rev. D}, 107, 2023{\natexlab{b}}.
\newblock \doi{10.1103/PhysRevD.107.076002}.

\bibitem[Ruffini et~al.(2010)Ruffini, Vereshchagin, and Xue]{Ruffini:2009hg}
R.~Ruffini, G.~Vereshchagin, and S.~S. Xue.
\newblock {Electron-positron pairs in physics and astrophysics: from heavy
  nuclei to black holes}.
\newblock \emph{Phys. Rept.}, 487:\penalty0 1--140, 2010.
\newblock \doi{10.1016/j.physrep.2009.10.004}.

\bibitem[Ruffini and Vereshchagin(2013)]{Ruffini:2012it}
R.~Ruffini and G.~Vereshchagin.
\newblock {Electron-positron plasma in GRBs and in cosmology}.
\newblock \emph{Nuovo Cim. C}, 036\penalty0 (s01):\penalty0 255--266, 2013.
\newblock \doi{10.1393/ncc/i2013-11500-0}.

\bibitem[Ferrer and Hackebill(2019)]{Ferrer:2019xlr}
E.~J. Ferrer and A.~Hackebill.
\newblock {Thermodynamics of Neutrons in a Magnetic Field and its Implications
  for Neutron Stars}.
\newblock \emph{Phys. Rev. C}, 99\penalty0 (6):\penalty0 065803, 2019.
\newblock \doi{10.1103/PhysRevC.99.065803}.

\bibitem[Ferrer and Hackebill(2023)]{Ferrer:2023pgq}
E.~J. Ferrer and A.~Hackebill.
\newblock {The Importance of the Pressure Anisotropy Induced by Strong Magnetic
  Fields on Neutron Star Physics}.
\newblock \emph{J. Phys. Conf. Ser.}, 2536\penalty0 (1):\penalty0 012007, 2023.
\newblock \doi{10.1088/1742-6596/2536/1/012007}.

\bibitem[Yan et~al.(2023)Yan, Ma, Ling, Cheng, and Huang]{Yan:2022sxd}
H.~Yan, Z.~Ma, C.~Ling, C.~Cheng, and J.~Huang.
\newblock {First Batch of z \ensuremath{\approx} 11\textendash{}20 Candidate
  Objects Revealed by the James Webb Space Telescope Early Release Observations
  on SMACS 0723-73}.
\newblock \emph{Astrophys. J. Lett.}, 942\penalty0 (1):\penalty0 L9, 2023.
\newblock \doi{10.3847/2041-8213/aca80c}.

\bibitem[Adams et~al.(2023)Adams, Conselice, Ferreira,
  et~al.]{adams2023discovery}
N.~J. Adams, C.~J. Conselice, L.~Ferreira, et~al.
\newblock Discovery and properties of ultra-high redshift galaxies $(9<z<12)$
  in the \uppercase{JWST} \uppercase{ERO} \uppercase{SMACS} 0723 field.
\newblock \emph{Monthly Notices of the Royal Astronomical Society},
  518\penalty0 (3):\penalty0 4755--4766, 2023.
\newblock \doi{10.1093/mnras/stac3347}.

\bibitem[Haro et~al.(2023)Haro, Dickinson, Finkelstein,
  et~al.]{arrabal2023spectroscopic}
P.~A. Haro, M.~Dickinson, S.~L. Finkelstein, et~al.
\newblock Confirmation and refutation of very luminous galaxies in the early
  universe.
\newblock \emph{Nature}, 2023.
\newblock \doi{10.1038/s41586-023-06521-7}.

\bibitem[Larson et~al.(2023)Larson, Finkelstein, Kocevski, Hutchison, Trump,
  Haro, Bromm, Cleri, Dickinson, Fujimoto, et~al.]{CEERSTeam:2023qgy}
R.~L. Larson, S.~L. Finkelstein, D.~D. Kocevski, T.~A. Hutchison, J.~R. Trump,
  P.~A. Haro, V.~Bromm, N.~J. Cleri, M.~Dickinson, S.~Fujimoto, et~al.
\newblock A \uppercase{CEERS} discovery of an accreting supermassive black hole
  570 myr after the big bang: Identifying a progenitor of massive $z>6$
  quasars.
\newblock \emph{arXiv preprint}, 2023.
\newblock \doi{10.48550/arXiv.2303.08918}.
\newblock [submitted to ApJ].

\bibitem[Jedamzik and Abel(2013)]{Jedamzik:2013gua}
K.~Jedamzik and T.~Abel.
\newblock {Small-scale primordial magnetic fields and anisotropies in the
  cosmic microwave background radiation}.
\newblock \emph{JCAP}, 10:\penalty0 050, 2013.
\newblock \doi{10.1088/1475-7516/2013/10/050}.

\end{thebibliography}

\end{document}